# Proximity and Josephson effects in microstructures based on multiband superconductors

Y. Yerin[a)]

*Institute for Physics of Microstructures, Russian Academy of Sciences, Nizhny Novgorod 603950, Russia and B.I. Verkin Institute for Low Temperature Physics and Engineering, National Academy of Sciences of Ukraine, 47 Nauki Ave., Kharkov 61103, Ukraine*

A. N. Omelyanchouk

*B.I. Verkin Institute for Low Temperature Physics and Engineering, National Academy of Sciences of Ukraine, 47 Nauki Ave., Kharkov 61103, Ukraine*

Emerging in the 1950s, the multiband superconductivity has been considered for a long time as an approximate model in the form of a generalization of the BCS theory to the case of two bands for a more accurate quantitative description of the properties and characteristics of such superconductors as cuprates, heavy fermions compounds, metal boron carbides, fullerides, strontium ruthenate etc. due to their complex piecewise-continuous Fermi surfaces. However the discovery of the multiband structure of the superconducting state in magnesium diboride in 2001 and iron oxypnictides and chalcogenides in 2008 led to the appearance of many papers in which effects and different dependences well known for conventional single-band s-wave superconductors were reexamined. The main purpose of these studies was to reveal the symmetry type of the order parameter, which provides an important information about the mechanism of Cooper pairing in these superconductors. One of the most effective methods of obtaining information on the symmetry properties of the order parameter in the multiband superconductors is phase-sensitive techniques. This review summarizes the results of theoretical and experimental studies of the proximity and Josephson effects in systems based on multiband superconductors in contact with normal metals, insulators and other superconductors.

## 1. Introduction

The microscopic theory of superconductivity by Bardeen, Cooper, and Schrieffer (BCS) is based on an isotropic metal model.[1,2] The results obtained in this approach agree qualitatively well with experiment, without, however, providing full quantitative agreement.

Almost immediately after the development of the BCS model, an attempt has been made to partially take into account the anisotropic properties of metals, namely, the effect of overlapping of the energy bands in the vicinity of the Fermi surface,[3,4] which leads to the appearance of interband quantum electron transitions and, as a result, to an additional indirect attraction between the electrons of each band . This gives rise to qualitatively new properties of the multiband superconductor model in comparison with the model of independent bands. It should be noted that at that time, the multiband (two-band) model of superconductivity was considered solely as an attempt to "fit" the experimental data obtained by studying various properties of single-band superconductors and later discovered compounds with so-called unusual superconductivity (cuprates, heavy-fermion compounds, borocarbides, fullerides, strontium ruthenate, organic superconductors) to the BCS theory.

However, the real boom in the study of multiband superconductivity began with the discovery of a superconducting transition in $MgB_2$ in 2001.[5] Using various experimental techniques, it was established that the superconducting phase of magnesium diboride has two independent gaps with different temperature dependences, the existence of which cannot be interpreted as the manifestation of gap anisotropy, as was usually the case for the so-called multiband superconductors mentioned above.[6–25]

In 2008, it was discovered that (oxy)pnictides and iron chalcogenides also exhibit a multiband superconducting state.[26] However, unlike $MgB_2$, the interaction between the order parameters characterizing the multi-gap nature of the superconductivity of these compounds is repulsive (the interband matrix elements of the interband interaction are negative). As a rule, the electronic band structure of iron-containing superconductors consists of two hole pockets in the center of the Brillouin zone and two electron pockets at $(\pi, \pi)$.[27–29] This must lead to the unique $s_\pm$-wave symmetry of the order parameter, which is remarkable due to isotropic superconducting gaps emerging on the hole and electron sheets of the Fermi surface, which, however, have opposite signs on these sheets.

It has been first proposed independently in Refs. 30–32 that such an order parameter is realized in iron-containing superconductors. It is fair to say that the concept of $s_\pm$ symmetry has been introduced even before the discovery of the superconducting phase in iron-based compounds in relation with the theoretical description of superconductivity in several other materials described by multiorbital models.[33–37]

Various electronic models of iron-containing superconductors, as well as the available experimental data,[38] also suggest that the $s_\pm$ superconducting state is nevertheless the most energetically optimal for this class of superconductors. Despite the rather large number of experimental facts supporting this hypothesis, the question of the symmetry of the order parameter, as well as the number of superconducting gaps in iron-containing superconductors, remains controversial, since for some members of the iron-based superconductor family, the two-band approach is not sufficient to provide full qualitative and quantitative explanation to the available data. Therefore, to better describe the experimental results, more complex chiral pairing symmetries have been proposed: $s + id$, with $s_\pm + is_{++}$ and isotropic s-wave tri-gap models.[39–42]

Obviously, the presence of a complex order parameter structure in multi-band superconductors generates an interesting new physics. In particular, such superconducting systems must lead to the emergence of a whole family of topological defects and quantum phenomena that have no analogs in conventional superconductors. For example, states with time-reversal symmetry breaking, collective modes of Leggett's type, phase solitons and domains, vortices carrying fractional magnetic flux and generating a unique intermediate state, different from both type-I and type-II superconductivity, as well as fractional Josephson effect.

Such a large number of new and non-trivial data have already been summarized in recent reviews devoted to the theoretical aspects of describing the topological defects of multiband superconductivity and their experimental detection.[43,44] Along with this, at the moment there are already a substantial number of papers addressing coherent current states in systems based on multiband superconductors, in particular, Josephson states, states in systems with doubly-connected geometry, and quasi-one-dimensional infinitely long channels.

It should be noted that, among other things, interest in the studies of the Josephson effect is due to the fact that phase-sensitive techniques often provide the most valuable information on the symmetry of the order parameter in unusual superconductors and are a sufficiently productive "tool" for identifying it in compounds where this problem remains unsolved (see, e.g., Refs. 45–48).

In view of the above, the present paper aims to provide an overview of the currently known theoretical and experimental results on current states in multiband superconductors and, in particular, the Josephson systems based on them.

The review is organized as follows: Section 2 addresses proximity effects that occur on the interface of a multiband superconductor with a normal metal or other superconducting material. The Josephson effect in multiband superconductors and the Josephson systems based on them are discussed in Sec. 3. The main results are summarized in Sec. 4. Also, for convenience, in Appendixes A and B, we provide a very brief overview of the microscopic theory of multiband superconductivity and the phenomenological Ginzburg-Landau model generalized to the case of multiple order parameters with s-wave symmetry.

## 2. Specific aspects of the proximity effect in heterostructures based on multiband superconductors

### 2.1. Contact of a normal metal and a two-band superconductor

The proximity effect is the phenomenon of the penetration of Cooper pairs of a superconductor into a normal metal or another superconductor over a distance of the order of the coherence length, leading to an induced energy gap in the material in contact with the superconductor. This effect has already been fairly well studied and described in terms of the Andreev reflection phenomenon and the microscopic formalism of Green's functions.[49–57]

Evidently, the presence of several energy gaps in the spectrum of quasiparticle excitations of a superconductor should impose certain features on the proximity effect. Such a problem has become topical, in particular, after the detection of multiband superconductivity in $MgB_2$, in oxypnictides and iron chalcogenides. The theoretical and experimental studies that appeared after that were primarily aimed at understanding how the multiband nature of the superconducting state influences the proximity effect, in particular, the density of states at the multiband superconductor–metal interface, tunnel current and other accompanying phenomena.

One of the first such studies theoretically investigated the proximity effect in bilayers consisting of a two-band superconductor and a single-band superconductor and those composed of a two-band superconductor and a normal metal.[58] In the paper, the formalism of the Usadel equations[59] and Kupriyanov-Lukichev boundary conditions[60,61] has been generalized to the case of two energy gaps (see Appendix A).

The first object for theoretical studies in this paper is a heterostructure formed by a single-band superconductor and $MgB_2$. Figure 1 shows the results of a numerical solution of the Usadel equations for energy gaps in a system formed by the superconducting magnesium diboride and a single-band conventional superconductor.

At temperatures above the critical temperature of the single-band superconductor (solid lines in Fig. 1), it can be seen that its gap grows as it approaches the interface, while $\Delta_\sigma$ (large gap in $MgB_2$) decreases, as, in principle, can be expected based on the analogy with the proximity effect in single-band superconducting bilayers. The decrease of the second, weaker gap $\Delta_\pi$ near the interface can be explained by a relatively strong interband coupling between the two bands. Due to the weakening of the interband interaction and the simultaneous increase in the pairing between the contacting superconductors (due to the small values of the barrier parameters at the heterostructure interface), an opposite regime has been achieved in which the energy gaps of the two-band superconductor grow toward the interface (dashed lines in Fig. 1). This result can be interpreted as a peculiarity of the proximity effect in the case of a two-band superconductor, in which the gap increases upon contacting a single-band superconducting material, which has a lower critical temperature.

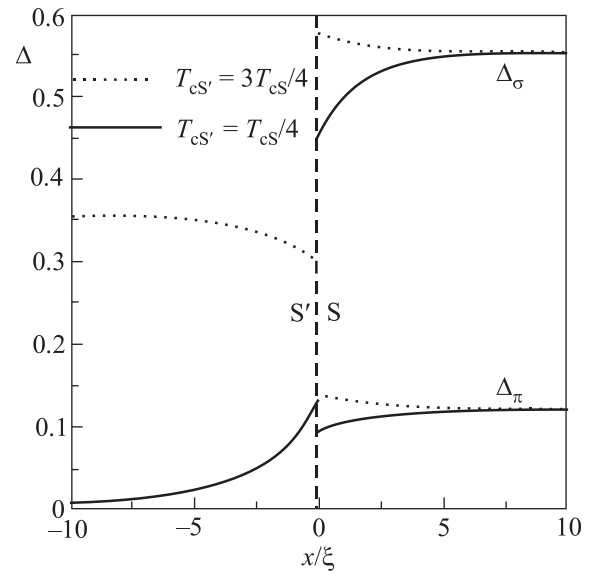

Fig. 1. Energy gaps of $MgB_2$ (S) and a single-band superconductor (S') as a function of the position in the S'S heterostructure. The coherence lengths of the superconductors are assumed identical. The dashed line denotes the bilayer interface.[58]

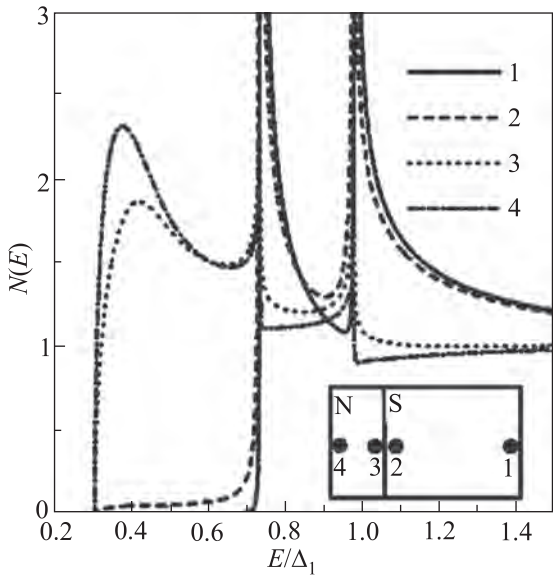

Fig. 2. Normalized density of states in an SN (S is the two-band superconductor, N is the normal metal) bilayer at four different points of the heterostructure (indicated by numbered black dots in the inset). The electron-phonon pairing constants for the two-band superconductor are $\Lambda_{11} = 0.5$, $\Lambda_{22} = 0.4$, $\Lambda_{12} = \Lambda_{21} = 0.1$.[58]

This behavior is confirmed by calculations of the density of states in an SN bilayer (Fig. 2). As follows from the graph, three peaks of the density of states are present in the normal layer of the heterostructure: the lowest layer corresponds to a small gap induced by the superconducting part of the bilayer, while the other two are the result of the two-band structure of the contacting superconductor.

### 2.2. An anomalous proximity effect at the boundary of s and $s_\pm$-wave superconductors

Superconducting magnesium diboride in an unperturbed state has a zero phase difference $\phi$ between the order parameters, or the $s_{++}$-wave pairing mechanism. The results of experiments with some compounds from the family of superconducting oxypnictides and iron chalcogenides imply that the $s_\pm$-wave symmetry of the order parameter in these superconductors is energetically favorable if $\phi = \pi$ in the ground state.

Obviously, the most convincing detection of the pairing mechanism is provided by phase-sensitive techniques. It is logical to assume that one of these tests to determine the opposite signs of order parameters in a two-band superconductor can be the proximity effect.

For the first time the possibility of using this effect has been theoretically substantiated in Ref. 62. For illustration, a "sandwich" consisting of a thin (compared to the coherence length) layer of an ordinary s-wave superconductor and a two-band superconducting material with $s_\pm$ type of symmetry has been considered under assumption that both substances have a high concentration of nonmagnetic impurities. The hypothesis of dirty superconductors allowed the authors to apply the formalism of the Usadel equations with the corresponding Kupriyanov-Lukichev boundary conditions. Within the framework of this approach and taking into account the assumed low thickness of the ordinary s-wave superconductor layer, the density of states on the interface of the system under study has been estimated analytically (Fig. 3).

Calculations have shown that the $s_\pm$-wave symmetry makes a unique "imprint" on the density of states of the heterostructure. The parallel orientation of the order parameter phases (see the inset in Fig. 3) leads to a minor peak in the density of states, while the antiparallel phases induce a small dip. While the presence of peaks in the density-of-states plot is the common effect between two superconductors, the anomalous dip observed in the calculations should obviously be considered as a feature arising due to the proximity to an $s_\pm$-wave two-band superconductor.

Further theoretical studies have shown that for certain characteristics of a bilayer formed by a conventional and an $s_\pm$-wave superconductors, a nontrivial frustrated state with broken symmetry with respect to time reversal can be formed in the heterostructure, corresponding to a phase difference different from zero or $\pi$.[63,64] The possibility of realizing this proximity effect can be demonstrated at a phenomenological level. The Josephson coupling energy for two superconductors in contact in the limit of weak coupling between the heterostructure layers has the form

$$E(\phi) = (E_{J1} - E_{J1})(1 - \cos\phi) + \frac{E_{J2}}{2}(1 - \cos 2\phi), \quad (1)$$

where $\phi$ denotes the phase difference between the order parameter $\Delta_s$ in a single-band superconductor and the first order parameter $\Delta_1$ in the $s_\pm$-wave two-band superconductor, and the coefficient $E_{J2}$ arises due to nonzero

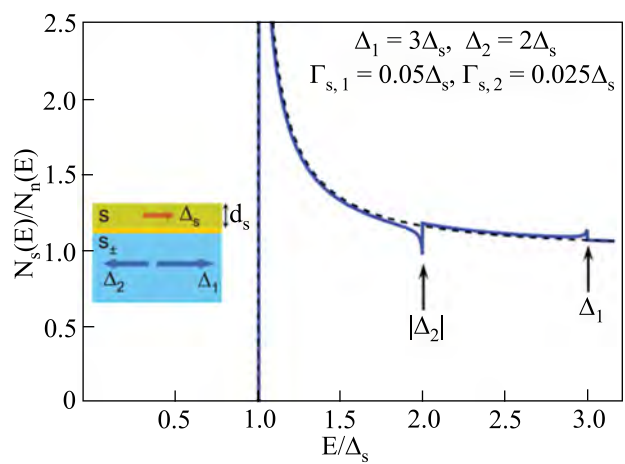

Fig. 3. The density of states at the interface of the heterostructure consisting of an ordinary s-wave superconductor and a two-band superconductor with $s_\pm$ symmetry type.[62]

transparency of the barrier at the bilayer interface. Expression (1) corresponds to the Josephson current:

$$j(\phi) = (j_{J1} - j_{J2})\sin\phi + j_J^{(2)}\sin 2\phi. \quad (2)$$

On the other hand, minimization of the energy (1) with respect to the phase difference $\phi$ gives an expression for the ground state of the heterostructure

$$\cos\phi_0 = \frac{j_{J1} - j_{J2}}{2|j_J^{(2)}|}. \quad (3)$$

Evidently, in the ground state, $\phi_0$ varies smoothly from 0 to $\pi$, while the current difference changes from $2|j_J^{(2)}|$ to $-2|j_J^{(2)}|$. Thus, in fact, the state with time-reversal symmetry breaking creates the so-called $\phi$-contact[65] (Fig. 4).

However, as the subsequent microscopic analysis showed, the second harmonic in the Josephson transport and, accordingly, the state with broken time-reversal symmetry arises only in the case of a very weak interband interaction in the two-band superconductor with the $s_\pm$-symmetry. As has been proposed in Ref. 64, both the presence of this state and the $s_\pm$-wave pairing mechanism can be determined using specific features in the density of states. The results of an approximate solution of the Usadel equations for the sandwich show the characteristic changes in the density of states (Fig. 5). In particular, the parallel orientation of the phases of the first order parameter of the two-band superconductor and the order parameter of the single-band superconductor ($\phi = 0$) generates a positive peak in the density of states, while the antiparallel orientation gives a negative contribution. In the "maximum" frustrated state, at $\phi = \pi/2$, according to calculations, the density of states has two small dips.

A more detailed study of the conditions for the appearance of the state with time-reversal symmetry breaking in the bilayer formed by a single-band superconductor and a two-band superconductor with $s_\pm$-wave symmetry has been carried out in Ref. 66. In the limit of low transparency of the barriers and at a temperature close to zero, the Usadel equations were solved analytically using the perturbation theory. Based on these results, a phase diagram was constructed, showing all possible states that arise due to the proximity effect between a single-band and two-band superconductors (Fig. 6) as a function of the barrier transparency at the interface.

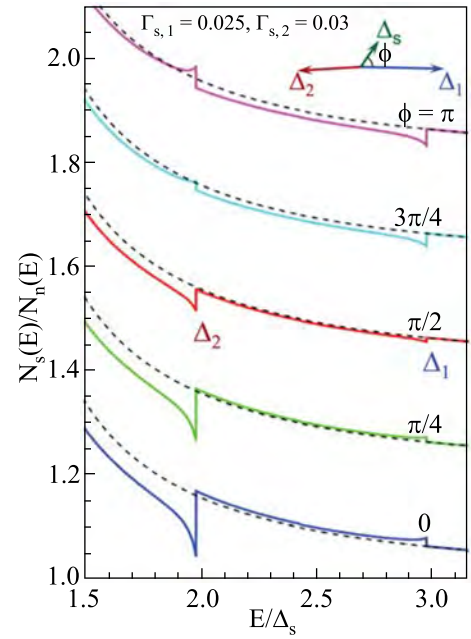

Fig. 5. Evolution of the singularities of the density of states in an s-wave superconductor, induced by the proximity effect with an $s_\pm$ superconductor. The difference between the first order parameter of the two-band superconductor and the order parameter of the single-band superconductor increases from 0 to $\pi$. For clarity, the curves are offset relative to each other. The dashed line corresponds to the density of states of an isolated massive single-band superconductor.[64]

Figure 6 illustrates the regions where the "positive" (the order parameter in the single-band s-wave superconductor increases) and "negative" (suppression of the order parameter in a single-band superconductor) proximity effects occur, as well as the small region of the so-called anomalous proximity effect, where the frustrated state is realized, leading, as was shown earlier, to breaking of time-reversal symmetry.

The proximity effect between a single-band and an $s_\pm$-wave superconductor can also be described in the framework of the Ginzburg-Landau phenomenological theory. In Ref. 67, the following Ginzburg-Landau free energy functional has been proposed to describe such a system

$$F[\Delta_1, \Delta_2] = F_L + F_R + F_c, \quad (4)$$

where $F_{v=L,R}$ is the energy of the left and right sides of the bilayer

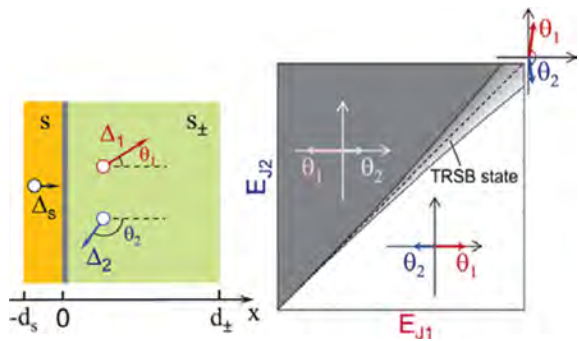

Fig. 4. (Left) Josephson junction of s and $s_\pm$ superconductors with arbitrary values of the order parameters. (Right) Schematic representation of the phase diagram of bilayer states as a function of the pairing energy of the order parameters of the two-band superconductor with the order parameter of the single-band superconductor.[64]

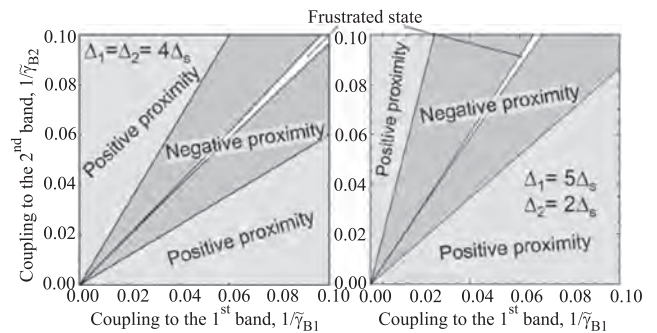

Fig. 6. Phase diagram of the possible states of a junction formed by s and $s_\pm$ superconductors in the weak-coupling limit between the heterocontact layers and at zero temperature for different values of the energy gaps shown in the figure.[66]

$$F_v = \int_v dx \left\{ \sum_{i=1,2} \left[ \frac{1}{2}\kappa_i^v |\partial_x \Delta_i|^2 - \frac{1}{2}r_i^v |\Delta_i|^2 + \frac{1}{4}u_i^v |\Delta_i|^2 \right. \right.$$
$$\left. \left. + \frac{1}{4}\alpha_i^v \left(\Delta_i^* \partial_x \Delta_i - \Delta_i \partial_x \Delta_i^*\right) \right] - v^v \left(\Delta_2^* \Delta_2 + \text{c.c}\right) \right\}. \quad (5)$$

$F_c$ represents the energy of a junction formed by a single-band and two-band superconductors

$$F_c = \sum_{i=1,2} \left[ T_i |\Delta_i(0^+) - \Delta_i(0^-)|^2 \right], \quad (6)$$

and $\kappa_i^v$, $r_i^v$ and $u_i^v$ denote the standard parameters of the phenomenological model. The term with $\alpha_i^v$ corresponds to an additional energy created by the superconducting current between the two superconductors. The coefficient $v^v$ describes the interband interaction. The coefficient $T_i$ is responsible for the transparency of the barriers at the heterostructure interface.

The variation of the free energy (4) with respect to the phase difference of the order parameter in a two-band superconductor with the $s_\pm$ type of symmetry makes it possible to classify possible regimes in bilayer behavior (Fig. 7).

Depending on the transparency of the barriers and the characteristics of the junction materials, there are three qualitatively different states: (1) the regime with a single minimum of the free-energy as a function of the phase difference of the order parameters of the two-band superconductor $\phi = 2\pi n$ ($n$ is an integer); (2) a state with broken time-reversal symmetry, where the free-energy minimum is degenerate for $\phi = 2\pi n \pm \phi_0$ and 3) a regime with two minima, one of which at $\phi = 2\pi n$ is global and another at $\phi = \pi(2n+1)$ is local.

Besides the formalism of the quasiclassical Usadel equations and phenomenology of the Ginzburg-Landau theory, Bogolyubov-de Gennes equations have also been used to study the characteristics of a junction and its probable singularities. In Ref. 68, they were used to develop the self-consistent theory of proximity effect in a heterostructure formed by an $s_\pm$-wave two-band superconductor and its single-band analogue. Numerical solution of the Bogolyubov-de Gennes equations allowed determine the energy of the system as a function of the phase difference between the order parameter of the single-band superconductor and the first order parameter of the $s_\pm$-wave two-band superconductor $E(\phi)$ (Fig. 8).

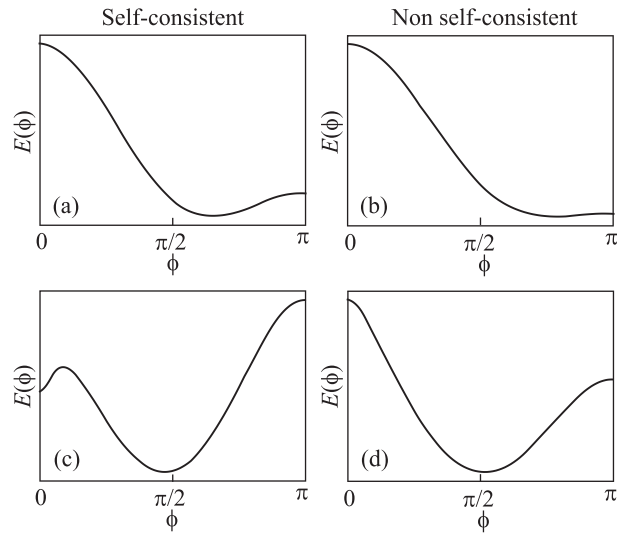

Fig. 8. Examples of the dependences of the energy on the phase difference between the order parameter in a single-band superconductor and the first order parameter in an $s_\pm$ wave two-band superconductor obtained from self-consistent calculations for a 0-contact (a) and a contact with two minima (c). Panels (b) and (d) show similar dependences obtained from non-self-consistent calculations for the same parameters as in panels (a) and (c), respectively.[68]

Since all the obtained dependences are periodic and have the symmetry $E(\phi) = E(-\phi)$, the graphs are plotted in the range 0 to $\pi$. Numerical simulation revealed four types of bilayer states: (a) 0-junction with the energy minimum at zero phase difference between the order parameter of the $s$-wave superconductor and that of the hole region of the Fermi surface of the two-band superconductor (accordingly, the phase difference between the order parameters of the conventional superconductor and that of the electron Fermi surface sheet of the two-band superconductor is $\pi$); (b) $\pi$-junction with the phase difference of $\pi$; (c) $\phi$-junction for the case of the energy minimum in the range $0 < \phi < \pi$; and (d) double-minimum junction with two minima on the $E(\phi)$ dependence, one is local, at zero phase difference, and another is global, located in the range $0 < \phi \leq \pi$.

Since the constructed model contains a rather large number of parameters of the sandwiched superconductors and the interface between them, it has been attempted (in the same paper) to find out the conditions for the formation of the four above states. To this end, the characteristics of the $s_\pm$-superconductor were fixed and the parameters of the interface and the conventional single-band superconductor were varied.

If the tunnel amplitudes of the bilayer, $w_x$ and $w_y$, are small, the transition between the 0-contact and the $\pi$-contacts is very sharp, and the region of existence of the $\phi$-contact occupies a very small "silver" region on the phase diagram (Fig. 9). With increasing $w_x$ and $w_y$, the phase space of the $\phi$-junction existence expands.

The chemical potential of the single-band superconductor $\mu_0$ also has a significant effect on the behavior of the heterostructure and its phase diagram. Figure 9 shows that there is a certain critical value of the chemical potential $\mu_c$, such that $\mu_0 > \mu_c > 0$, for which the energy minimum of the system is realized for the phase difference equal to $\pi$. For $\mu_0 < \mu_c$, small changes in the chemical potential lead to transitions from the zero phase difference to $\pi$ and vice versa. It was also found that increasing the gap in the single-band superconductor reduces the number of these transitions at fixed values of the tunnel amplitudes.

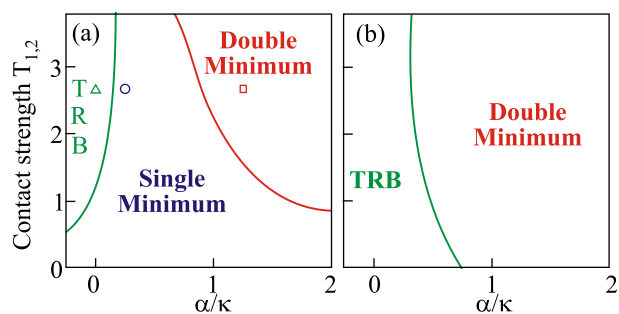

Fig. 7. (a) Phase diagram of the possible states of an s–$s_\pm$ bilayer for the following coefficients of the Ginzburg-Landau theory: $r_i^v = 1$, $\kappa_i^v = 4$, $u_1^v = 1$, $u_2^v = 2$, $v^L = -v^R$, $\alpha_i^v = \alpha$, $l = 3$, and $T_1 = T_2$. Different states correspond to qualitatively different behavior of the dependences of the energy on the phase difference of the order parameters in the two-band superconductor: a state with a single minimum ("single minimum") at $\phi = 2\pi n$, a phase with time-reversal symmetry breaking ("TRB") with a minimum at the points $\phi = 2\pi n \pm \phi_0$ and a state with two minima ("double minimum") at $\phi = \pi n$. (b) The same as in panel (a), but with the identical order parameter moduli in the two-band superconductor. In this case, the state with a single minimum disappears.[67]

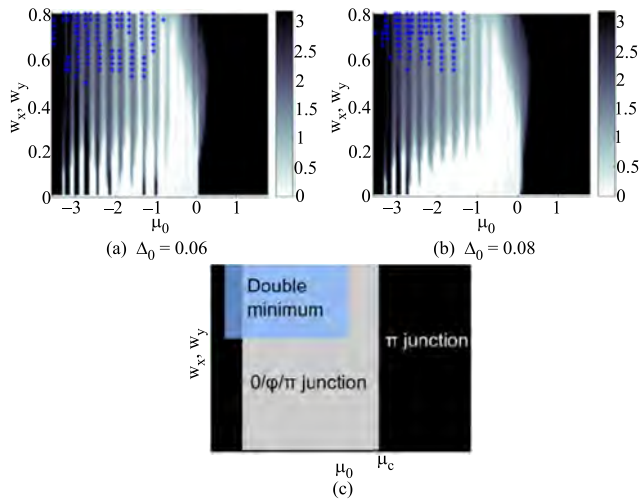

Fig. 9. Phase diagrams, (a) and (b), showing the states of the s–s$_\pm$ superconductor heterostructure as a function of the tunnel amplitudes $w_x = w_y$ and the chemical potential $\mu_0$ of the single-band superconductor with a gap $\Delta_0$. The white regions correspond to zero phase difference between the s-wave superconductor and the order parameter for the electron pocket, while the black regions correspond to the phase difference of $\pi$. Gray regions indicate an intermediate state with a phase difference from 0 to $\pi$. The blue symbols indicate regions with two minima on the energy dependence on the phase difference. Panel (c) shows schematics of the phase diagram. In black regions, the energy of the system is reached at the phase difference of $\pi$, in gray regions the entanglement of states with a phase difference of 0, $\varphi$ and $\pi$ is realized. The region in which an additional minimum with zero phase difference can possible exists is marked in blue.[68]

In addition to the fundamental aspects outlined above, interest in the proximity effect between a normal metal and a multi-band superconductor is also due to the fact that it is the theoretical basis for microscopic Andreev spectroscopy, which serves as a powerful method for diagnosing the symmetry of the order parameter in superconducting materials with several energy gaps.[69–71] In Ref. 71, the Blonder-Tinkham-Klapwijk formalism was generalized to the case of a two-band superconductor, taking into account the phase difference of the order parameter in it. The analytical expression for the conductivity of the microcontact obtained from the solution of the Bogolyubov-de Gennes equations made it possible to establish an important difference between s$_{++}$ and s$_\pm$-wave symmetry in a two-band superconductor, which can be used to interpret the corresponding experiments of microcontact spectroscopy of "iron" superconductors.[72–76]

## 3. Josephson effect in multiband superconductors

In 1962, Josephson predicted that persistent current can flow through a tunnel junction formed by two superconductors. It has been shown that the magnitude of this current depends sinusoidally on the coherent phase difference between the left and right superconducting banks.[77] This effect was later called the dc Josephson effect.

In Ref. 78, Josephson established that when the tunnel current exceeds a certain critical value, a non-zero voltage appears across the junction, and the current becomes non-stationary oscillating with a frequency proportional to the voltage. This phenomenon is called the ac Josephson effect. These theoretical predictions were soon confirmed experimentally.[79–81]

The detection of two-band superconductivity in magnesium diboride and the multi-band structure of the energy spectrum of the superconducting state in oxypnictides and iron chalcogenides revived the interest in phase-coherent current states, among which the Josephson effect also plays an important role. Taking into account the fact that high-temperature iron-based superconductors probably possess complex symmetry of the type s$_\pm$ s + $id$, s$_\pm$ + $is_{++}$, etc., Josephson spectroscopy is in fact the only method that makes it possible to "sense" the phase difference between the order parameters in these compounds.

At the same time, the Josephson effect is also of use for detecting collective modes in multiband superconductors, first predicted by Leggett.[82] These modes can exist independently of the signs of the order parameters in the bands and include oscillations of the phase difference.

### 3.1. SNS and SIS junctions

The Josephson effect can be realized in different structures.[83] In Josephson's pioneering work, a tunnel junction with a delta-barrier was considered. Further theoretical studies have shown that a nondissipative flow of current is also possible in SIS tunnel junctions with weak coupling in the form of an insulating (I) interlayer between two superconductors (S), as well as SNS sandwiches with a layer of normal metal (N).

After the discovery of superconductivity in magnesium diboride, it became clear that, being a phase-sensitive phenomenon, this effect should have unique characteristics in multiband superconductors that reflect the unusual symmetry of the order parameter. The first works in search of these features were devoted to the SIS and SNS Josephson junctions, where one or both superconductors have s$_\pm$ symmetry.

Within the framework of the quasiclassical Eilenberger equations and the respective generalization of the Kupriyanov-Lukichev boundary conditions, one can derive the Ambegaokar-Baratov equation for the junctions formed by multiband (two-band) superconductors[84] (a more detailed analysis of the equation for iron superconductors has been carried out in Ref. 85).

$$I_{ij} = \frac{\pi T}{e R_{ij}} \sum_\omega \frac{\Delta_{Rj}\Delta_{Li}}{\sqrt{\omega^2 + \Delta_{Li}^2}\sqrt{\omega^2 + \Delta_{Rj}^2}}, \quad (7)$$

where $L$ and $R$ denote the left and right superconductors, respectively, and $R_{ij}^{-1} = \min\{R_{Lij}^{-1}, R_{ij}^{-1}\}$ is the normal junction resistance for the bands $(i, j)$, defined by a special integral over the Fermi surface $S_{Li(Rj)}$

$$(R_{L(R)ij}A)^{-1} = \frac{2e^2}{\hbar} \int_{v_x>0} \frac{D_{ij} v_{n,Li(Rj)} d^2 S_{Li(Rj)}}{(2\pi)^3 v_{F,Li(Rj)}}. \quad (8)$$

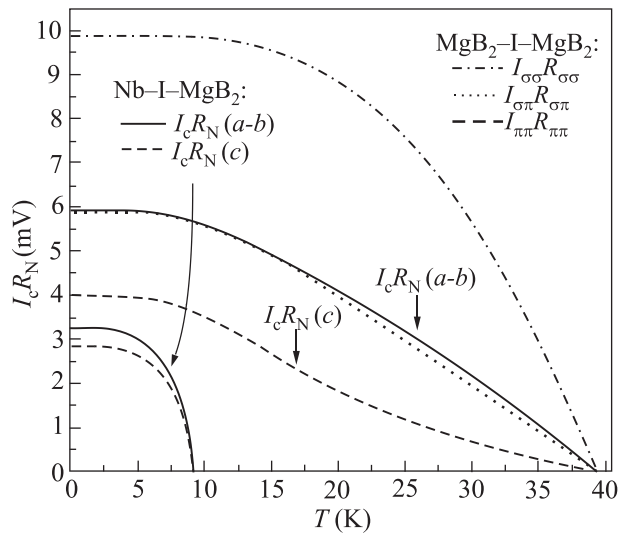

Fig. 10. Temperature dependences of the critical currents of tunnel junctions $MgB_2$-I-$MgB_2$ and Nb-I-$MgB_2$ for different bands and different crystallographic orientations.[84]

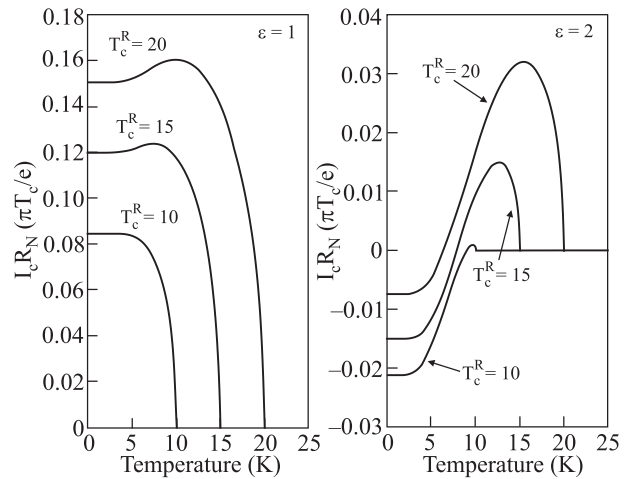

Fig. 11. Temperature dependence of the Josephson current in the junction formed by an $s_\pm$ two-band superconductor with a critical temperature $T_c$ = 41 K and a single-band superconductor with different $T_c^R$ shown in the graph. The parameter $\varepsilon = R_{N,\sigma}/R_{N,\pi}$ reflects the partial contributions of the $\sigma$ and $\pi$ bands of the superconductor with two gaps to the resistance of the tunnel junction. The coupling constants in a two-band superconductor are $V_{\pi\pi} = 0$, $V_{\sigma\pi} = V_{\pi\sigma} = -0.35|V_{\sigma\sigma}|$.[86]

Here $A$ is the junction, $v_n$ is the projection of the Fermi velocity $v_F$ on the normal direction to the interface plane, $D_{ij}$ denotes the tunneling probability of a quasiparticle from the band $i$ of the left superconductor $L$ to the band $j$ of the right bank $R$. The total current $I$ through the contact is obtained by summing up $I = \sum_{ij} I_{ij}$.

Almost immediately after the electron-phonon mechanism of two-band superconductivity in $MgB_2$ was established and the coupling constant matrices in this compound were determined, this equation was applied to the calculating of the Josephson tunneling in $S_2IS_2$ and $S_1IS_2$ contacts, where $S_1$ and $S_2$ are single-gap niobium and double-gap magnesium diboride, respectively.[84] The results of numerical calculations are shown in Fig. 10.

Owing to strong intra- and interband pairing, the temperature dependence of the critical current of the $MgB_2$-I-$MgB_2$ system for the $ab$ crystallographic plane differs from that for a single-band isotropic superconductor obtained from the Ambegaokar-Baratoff relation. Due to the predominant contribution from the $\pi$-band, the critical current in the $ab$ plane shows a positive curvature. For tunneling along the $c$-axis, the critical current is formed solely by the contribution from the $\pi$-band.

A similar behavior of the heterostructure is realized also for the case when one of the junction banks is a single-band superconductor Nb-I-$MgB_2$. The critical current of the system along the $c$-axis is created only by the $\pi$-band.

The most interesting result is produced by the Ambegaokar-Baratoff relation in the case of a tunnel current in a Josephson system in which one of the electrodes is a two-band superconductor with a phase difference in the banks equal to $\pi$, i.e., with $s_\pm$ symmetry, and the other electrode is a conventional superconductor with a single gap. It has been shown in Ref. 86 that in the case of equal partial contributions to the normal junction resistance, the temperature dependence of the critical current through the heterostructure exhibits a maximum that does not coincide with absolute zero (Fig. 11), as would be the case for the Josephson effect in single-band superconductors.

If the contributions of each band to the normal resistance are not equal, the behavior of the temperature dependence becomes even more exotic. At a certain temperature different from the superconducting transition temperature, the critical current of the contact is zero (Fig. 11). In fact, this means that an increase in temperature shifts the ground state of the Josephson system from 0 to $\pi$, i.e., there is temperature-induced switching from a normal contact to $\pi$-contact.

A deeper analysis of the 0–$\pi$ switching conditions for the ground state in ballistic SIS heterostructures has been carried out within the Bogolyubov-de Gennes equations.[87] In this paper, a contact formed by s and $s_\pm$ superconductors has been considered. Using numerical simulation, the spectra of Andreev bound states were calculated as a function of the interband interaction coefficient in a two-band superconductor for coinciding energy gaps and zero temperature. While the intersection of the energy states (blue curve) is observed for the unpaired non-interacting bands, with increasing the "force" of interband interaction, they start to "repulse" each other, and a growing gap is formed in the spectrum of the Andreev states [Fig. 12(a)].

However, one of the main results of Ref. 87 is the possibility of 0–$\pi$ switching in an SIS system with a certain ratio of transparency coefficients for each band on the interface of the Josephson junction. It manifests itself as a salient point in the dependence of the critical current of the system [Fig. 12(b)].

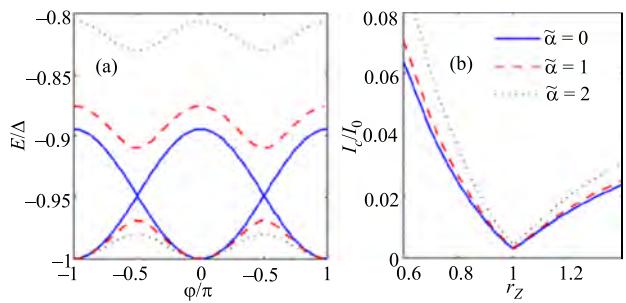

Fig. 12. Andreev bound states (a) and the critical current $s|I|s_\pm$ of the junction as a function of the ratio of the transparency coefficients of the barriers $r_Z$ at the interface of the system (b) for different values of the interband interaction $\tilde{\alpha}$ in the two-band superconductor.[87]

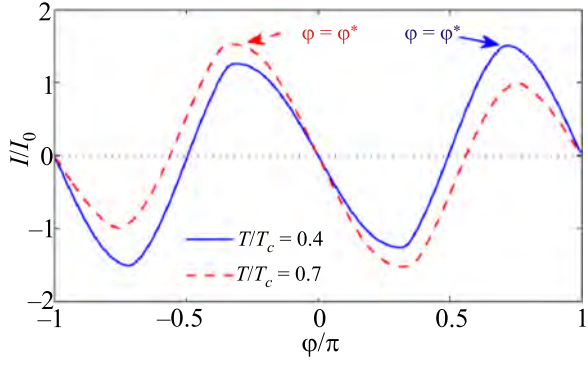

Fig. 13. Current-phase relationships for the s|I|$s_\pm$ contact at $T/T_c = 0.4$ (blue curve) and $T/T_c = 0.7$ (red dashed curve) for the ratio of energy gaps in a two-band superconductor equal to 0.3 and the ratio of the energy gap in a single-band superconductor and the first gap of a two-band gap of 0.5. The arrows indicate the position of the maximum of the Josephson current.[87]

For nonzero temperature and mismatching values of the energy gaps, the current-phase relations exhibit $\pi$-junction behavior with the dominant component due to the second harmonic (Fig. 13). This gives the maximum of the Josephson current at a certain phase difference $\varphi^*$ different from $\pi/2$.

The same results, in particular, the temperature-induced crossover from the conventional to $\pi$-junction behavior and the appearance of the second-harmonic component in the Josephson transport, can be obtained by applying the method of tunneling Hamiltonian.[88]

The diversity of the ground states of $s_{++}$I$s$ and $s_\pm$I$s$ junctions has also been demonstrated using the Gor'kov equations for the Green's functions in the approximation of strongly coupled electrons[89,90] with various orientations of the interface with respect to the crystallographic axes of two-band superconductors taken into account. For certain specific parameters, corresponding to real microscopic parameters of some iron-based superconductors, the current-phase dependences of $s_\pm$I$s_\pm$ heterostructures exhibit a $\pi$-contact signature.

Iron superconductors, some of which are unique in their $s_\pm$ type symmetry, are also noteworthy since their temperature-doping phase diagram contains a region where the superconducting dome intersects with the magnetic phase characterized by spin-density waves (SDW). The effect of SDW on the characteristics of $s_\pm$I$s_{++}$, $s_{++}$I$s_\pm$ and $s_\pm$I$s_\pm$ junctions has first been analyzed in Ref. 91. It turned out that for an $s_\pm$I$s_\pm$ heterostructure, its critical current contains a term that depends on the orientation angle of the magnetization vector in the SDW state. However, both in the presence and absence of SDW in two-band superconductors, there are conditions under which a $\pi$-junction in the $s_\pm$I$s_\pm$ system can be observed in appropriate experiments with superconducting oxypnictides and iron chalcogenides.

Replacement of the insulating interlayer in the junction by a normal metal does not qualitatively change the specific features of the Josephson effect in an $s_\pm$ superconductor. The numerical solution of the Usadel equations with the generalized Kupriyanov-Lukichev boundary conditions for the case of a two-gap superconductor has shown that the ratio of the barrier transparency coefficients for the interfaces of each band $r_\lambda$ ($\lambda = 1,2$) in both SNS and SIS junctions plays a key role in the realization of the 0–$\pi$ transition.[92,93] In other words, if a two-band superconductor has $s_\pm$ symmetry, then for a certain value of $r_\lambda$, the critical current of the contact exhibits a kink in the $I_c(r_\lambda)$ dependence, which corresponds to the crossover from the conventional junction to $\pi$-junction (Fig. 14).

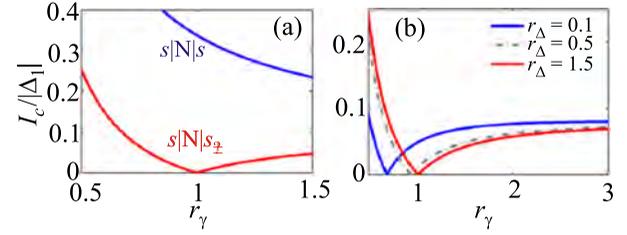

Fig. 14. Dependence of the critical current of s|N|s and s|N|$s_\pm$ junctions on the ratio of the transparency coefficients of the barriers at the interface for each of the bands $r_\gamma$, plotted for the ratio of the energy gaps of the two-band superconductor $r_\Delta = |\Delta_2|/|\Delta_1| = 1$ and the ratio of the energy gap of the single-band superconductor to the first energy gap of the $s_\pm$ superconducting layer equal to 1 (a) and 0.5 (b). For both graphs, the ratio of the thickness of the normal metal layer to the coherence length in the single-band superconductor was assumed 1.[92]

According to the calculations, a similar kink also occurs in the temperature dependence of the critical current of an SNS heterostructure for certain parameters of the contacting superconductors, in which (as above) one of the superconducting banks has two order parameters with a phase difference of $\pi$ in the ground state (Fig. 15).

A more general theory of ballistic Josephson junctions where both banks are identical multiband superconductors with $s_{++}$ or $s\pm$ symmetry has been developed in Refs. 94 and 95 The model is based on the Bogolyubov-de Gennes formalism, and the Blonder-Tinkham-Klapwijk potential is employed as an interface barrier, which allows to describe both the SIS and SNS junction types. Numerical solution of the equations allowed to obtain the spectrum of the Andreev bound states of the heterostructure, from which it is possible to directly extract the current-voltage dependences (Fig. 16).

From Fig. 16, several conclusions can be drawn. First, the Josephson junction formed by $s_{++}$ superconductors exhibits higher critical current. Second, and most importantly, only in the case of superconductors with $s_\pm$ wave symmetry and only for certain ratios of the parameters of the interface and superconducting banks, it is possible to realize a $\pi$-junction.

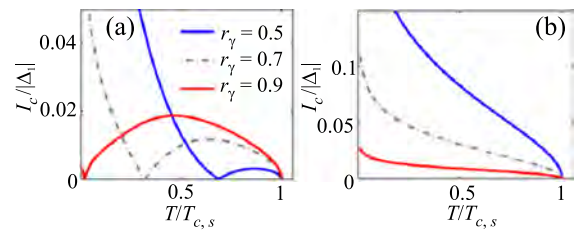

Fig. 15. Critical current of an s|N|$s_\pm$ junction as a function of the temperature, plotted for the ratio of the energy gap of the single-band superconductor to the first energy gap of the $s_\pm$ superconducting layer equal to 0.5 and the ratio of the thickness of the normal metal layer to the coherence length in the single-band superconductor equal to 1 for different values of $r_\gamma$ (see Fig. 14). For the graph in panel (a), $r_\Delta = 0.3$; for panel (b) $r_\Delta = 1.3$.[92]

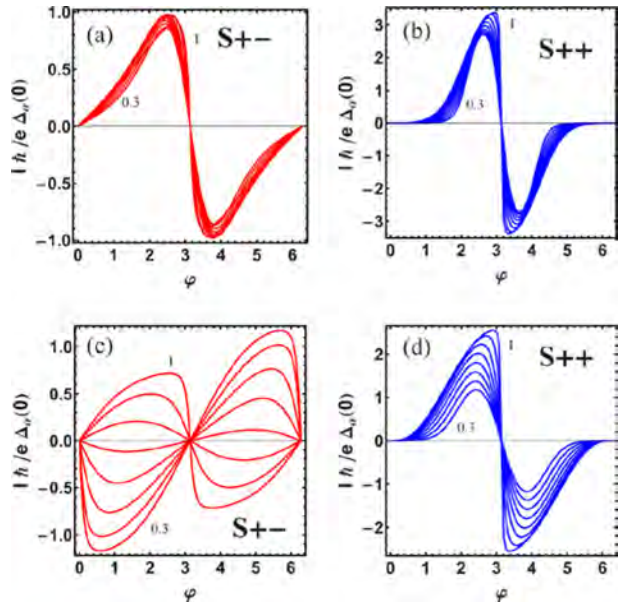

Fig. 16. Current-phase dependences for Josephson junctions formed by $s_\pm$ (a) and (c) and two-band superconductors with $s_{++}$ (b) and (d) symmetry for different interface parameters of the heterostructure and superconducting banks. Temperature $T = 0.001 T_c$.[95]

The temperature behavior of the critical current further confirms that the ground state of a Josephson system with a phase difference of $\pi$ can be realized (Fig. 17). Furthermore, it follows that for both types of symmetries, $s_{++}$ and $s_\pm$, the critical current deviates from the temperature behavior described by the Ambegaokar-Baratoff relation for the heterostructures formed by single-band superconductors.

In the previous section on the proximity effect, it was pointed out that in the bilayers with an $s_\pm$ superconductor it is possible to realize a state with broken time-reversal symmetry. A similar state can be also observed under certain conditions in the Josephson junctions in which one of the layers is a two-band superconductor of the $s_\pm$ type, and the other is a conventional s-wave superconductor. In Ref. 96, the conditions under which it is energetically favorable for a Josephson system to acquire a state with broken time-reversal symmetry have been found phenomenologically. This can be shown by varying the junction energy in the absence of a magnetic field and transport current in terms of $\phi_{s1}$ and $\phi_{s2}$ (the phase differences between the first/second order parameter of a two-band superconductor and the order parameter of its single-band analogue):

$$E = -\cos\phi_{s1} - J_{s2}\cos\phi_{s1} - J_{12}\cos(\phi_{s1} - \phi_{s2}), \quad (9)$$

where $J_{s2}$ is the energy of the Josephson coupling between the s superconductor and the second band of the $s_\pm$ superconductor, and $J_{12}$ is the energy of the interband interaction in the $s_\pm$ superconducting part of the system (for convenience, both quantities are normalized to the energy of the Josephson coupling between the single-band superconductor and the first band of the two-band superconductor $J_{s1}$).

The result of minimizing energy Eq. (9) is shown in Fig. 18(a), where the region of the phase diagram with broken time-reversal symmetry is visible, as well as in Fig. 18(b) showing the energy landscape of all possible states of the Josephson junction. As follows from the phase diagram, time-reversal symmetry breaking is observed mainly at $J_{12} < 0$, when the two-band superconductor has the $s_\pm$ order-parameter structure. It should be noted that the same conclusion have been obtained earlier for bilayer systems (see Sec. 2.2).

In addition to the above-mentioned features such as a kink in the temperature dependence of the critical current, 0–$\pi$ transitions, unusual current-phase dependences and states with broken time-reversal symmetry, the $s_\pm$-wave symmetry of the order parameter also affects the ac Josephson effect, in particular the microwave response of a Josephson junction. In several papers,[96,97] it has been shown theoretically the current-voltage characteristic of a heterostructure formed by s and $s_\pm$ superconductors, exhibit additional current jumps in addition to the conventional Shapiro steps. In this case, the step heights at the resonance voltage frequency demonstrate an alternating structure, which corresponds to the presence of two gaps in the superconductor. Moreover, as shown in Ref. 97, in the case of $s_{\pm}$ symmetry, under certain conditions, there is a significant enhancement of steps with odd multiplicity with respect to the resonant frequency of external microwave irradiation, while in the case of $s_{++}$ two-band superconductor, steps with even multiplicity dominate.

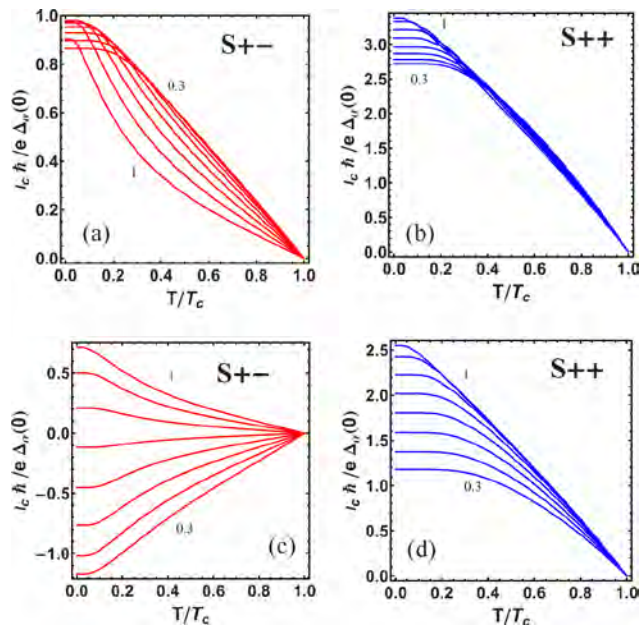

Fig. 17. Temperature dependence of the critical current of the Josephson junctions formed by $s_\pm$ (a) and (c) and two-band superconductors with $s_{++}$ (b) and (d) symmetry for different parameters of the heterostructure interface and superconducting banks.[95]

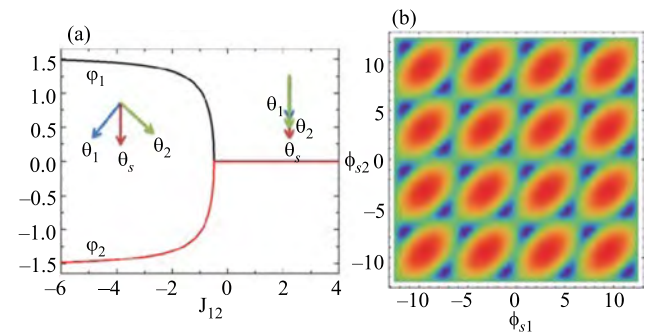

Fig. 18. (a) Phase diagram of the Josephson junction: the dependence of the phase difference $\phi_{s1}$ and $\phi_{s2}$ on the energy of the interband interaction in an $s_\pm$-wave superconductor with $J_{s1} = J_{s2} = 1$. The region with noncollinear "vectors" $\theta_s$, $\theta_1$, and $\theta_2$, bounded by the black and red curves, the state of the system with time-reversal symmetry breaking (the twofold degeneracy of the ground state). (b) The contour map of the energy of the Josephson junction $E = E(\phi_{s1}, \phi_{s2})$ for $J_{12} = -1$ and $J_{s2} = 1$. The blue regions correspond to the energy minima.[96]

Two-band $s_{\pm}$ superconductivity also generates unique signatures of topological excitations in the contacts, which, unlike fractional vortices in massive multi-band superconductors, possess a finite energy and, as a consequence, are thermodynamically metastable formations. We are talking about Josephson vortices and collective modes, the structure of which reflects the $s_{\pm}$ wave symmetry.[98–100] A detailed description of the characteristics and properties of these topological defects in multi-band superconductors can be found in the reviews Refs. 43 and 44.

### 3.2. ScS microcontacts

An ScS microcontacts, where the letter "c" denotes a constriction, is the simplest system for the experimental observation of the Josephson effect. Depending on the relation between the constriction size $d$ and the mean free path $l$, there are two types of contact: the Sharvin microcontact in the case of $d \ll l$,[101] and the diffusion microcontact, also known as the Maxwell microcontact with $d \gg l$. A so-called quantum microcontact deserves a separate mention: its constriction size is comparable with the Fermi wavelength. In this chapter, we imply the Sharvin microcontact.

A Josephson system with a constriction is formed by placing a superconducting needle on the surface of a massive superconductor or by depositing a superconductor onto a bilayer consisting of an insulator and a superconductor with a microscopic hole made in the insulator.

From the theoretical point of view, a microcontact is usually modeled as a quasi-one-dimensional thread connecting two massive banks. In the diffusion limit for single-band superconductors, the Josephson current can be calculated from the Usadel equations, in which all terms but the gradient ones are neglected. For symmetric contact (identical superconductors), the calculation results are known in the literature as the Kulik-Omelyanchuk theory, or, following the terminology of Ref. 83, the KO-1 theory.[102] The main conclusion of the KO-1 theory is the non-sinusoidal form of the current-phase dependence of a microcontact at zero temperature, which, as the temperature increases, evolves into a sinusoidal curve described by the Aslamazov-Larkin model.[103] A generalization of the KO-1 theory to the case of different contacting superconductors has been carried out in Ref. 104. It has been shown that there exists a logarithmic crossover to the sinusoidal form of the current-phase dependence in the limit of an infinitely large ratio of the energy gaps of superconducting banks.

In the pure limit, the Kulik-Omelyanchuk (KO-2) theory has been developed, in which the current-phase dependence of a Josephson system with ballistic conductivity is calculated within the framework of the Eilenberger equations.[105] The characteristics of a microcontact with an arbitrary

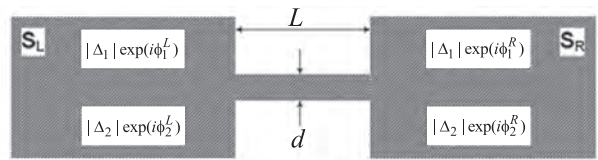

Fig. 19. ScS contact model. The right and left banks are massive two-band superconductors connected by a thin superconducting filament of length $L$ and diameter $d$. The relations $d \ll L$ and $d \ll \min[\xi_1, \xi_2]$ are satisfied.[108]

barrier transparency in the ScS junction have been obtained in Refs. 106 and 107.

The Josephson effect in microcontacts formed by "dirty" two-band superconductors has first been considered in Ref. 108 within the framework of the phenomenological Ginzburg-Landau theory. Due to the system geometry (Fig. 19), in the two-component Ginzburg-Landau equations all terms but the gradient terms are neglected. This approach allows to obtain an analytic expression for the Josephson current:

$$j \equiv j_0 \sin \phi$$
$$= \frac{2e\hbar}{L} \left( \frac{|\Delta_{01}|^2}{m_1} + \frac{|\Delta_{02}|^2}{m_2} + 4\,\text{sgn}\,(\gamma)\eta|\Delta_{01}||\Delta_{02}| \right) \sin \phi, \quad (10)$$

where $\phi \equiv \phi_1^R - \phi_1^L = \phi_2^R - \phi_2^L$, and the phenomenological constants $\gamma$ and $\eta$ correspond to the effects of interband interaction and interband scattering in a two-band superconductor.

It follows from Eq. (10) that the quantity $j_0$ can take both positive and negative values, provided that

$$j_0 > 0 \quad \text{for } \eta \,\text{sgn}\,(\gamma) > -\left(\frac{1}{4m_1}\frac{|\Delta_{01}|}{|\Delta_{02}|} + \frac{1}{4m_2}\frac{|\Delta_{02}|}{|\Delta_{01}|}\right), \quad (11)$$

$$j_0 < 0 \quad \text{for } \eta \,\text{sgn}\,(\gamma) < -\left(\frac{1}{4m_1}\frac{|\Delta_{01}|}{|\Delta_{02}|} + \frac{1}{4m_2}\frac{|\Delta_{02}|}{|\Delta_{01}|}\right). \quad (12)$$

If condition (12) is satisfied for a given set of parameters of a two-band superconductor, then the microbridge can behave as a $\pi$-contact.

It should be noted that later in Ref. 109, the criterion for the absolute minimum of the Ginzburg-Landau functional has been established (see Appendix B), which imposes a restriction on the value of the coefficient $\eta$

$$|\eta| < \frac{1}{2\sqrt{m_1 m_2}}. \quad (13)$$

Taking into account inequality (13), the quadratic form (12) is positive definite for any values of $|\Delta_{01}|$, $|\Delta_{02}|$, $\gamma$ and admissible values of $\eta$, since

$$\frac{|\Delta_{01}|^2}{m_1} + \frac{|\Delta_{02}|^2}{m_2} + 4\eta|\Delta_{01}||\Delta_{02}|\,\text{sgn}(\gamma)$$
$$> \frac{|\Delta_{01}|^2}{m_1} + \frac{|\Delta_{02}|^2}{m_2} - \frac{2|\Delta_{01}||\Delta_{02}|}{\sqrt{m_1 m_2}}\,\text{sgn}(\gamma)$$
$$= \left(\frac{|\Delta_{01}|}{m_1} - \frac{|\Delta_{02}|}{m_2}\,\text{sgn}(\gamma)\right)^2 \geq 0, \quad (14)$$

therefore, within the framework of the phenomenological description, the existence of $\pi$-contact is impossible.

Nevertheless, the current-phase dependence (10) remains valid for two-band superconductors with very weak interband scattering when $\eta \ll \frac{1}{2\sqrt{m_1 m_2}}$.

A microscopic analysis of the characteristics of the ScS system (Fig. 19), formed by dirty multiband superconductors, can be carried out using a suitable generalization of the KO-1 theory. For two-band superconductors this has been achieved in Refs. 110 and 111. As in the single-band case, the one-dimensional Usadel equations (see Appendix A) allows an analytic solution, which yields the current density in the form

$$j = \frac{4e\pi T}{L}\sum_i \sum_\omega^{\omega_D} N_i D_i \frac{|f_i|\cos\frac{\chi_i^L - \chi_i^R}{2}}{\sqrt{(1-|f_i|^2)\sin^2\frac{\chi_i^L - \chi_i^R}{2} + \cos^2\frac{\chi_i^L - \chi_i^R}{2}}}$$

$$\times \arctan \frac{|f_i|\sin\frac{\chi_i^L - \chi_i^R}{2}}{\sqrt{(1-|f_i|^2)\sin^2\frac{\chi_i^L - \chi_i^R}{2} + \cos^2\frac{\chi_i^L - \chi_i^R}{2}}}, \quad (15)$$

where $f_i(x) = |f_i(x)|\exp(i\chi_i(x))$ represent the anomalous Green's functions for each band, $D_i$ are the intraband diffusion coefficients due to the presence of scattering on nonmagnetic impurities, $N_i$ is the density of states on the Fermi surface for the $i$-th band electrons.

This general expression describes the Josephson current as a function of the energy gaps in the banks $|\Delta_i|$ and the phase difference at the junction $\phi \equiv \phi_1^R - \phi_1^L = \phi_2^R - \phi_2^L$. If the interband scattering effect is neglected, the current density (15) consists of two independent additive contributions arising from $1 \rightarrow 1$ and $2 \rightarrow 2$ transitions. In this case, each component of the Josephson current is proportional to the respective values $|\Delta_{1,2}|$.

$$I = \frac{2\pi T}{e}\sum_i \sum_\omega \frac{1}{R_{Ni}} \frac{|\Delta_i|\cos\frac{\phi}{2}}{\sqrt{\omega^2 + |\Delta_i|^2 \cos^2\frac{\phi}{2}}}$$

$$\times \arctan \frac{|\Delta_i|\sin\frac{\phi}{2}}{\sqrt{\omega^2 + |\Delta_i|^2 \cos^2\frac{\phi}{2}}}, \quad (16)$$

where $R_{Ni}$ are the partial contributions to the Sharvin resistance of the microcontact $R_N$.

At $T = 0$, it is possible to go in Eq. (16) from summation to integration over the Matsubara frequencies and substantially simplify the form of the current-phase dependence

$$I = \frac{\pi|\Delta_1|}{2eR_{N1}}\cos\frac{\phi}{2}\,\text{Artanh}\sin\frac{\phi}{2} + \frac{\pi|\Delta_2|}{2eR_{N2}}\cos\frac{\phi}{2}\,\text{Artanh}\sin\frac{\phi}{2}. \quad (17)$$

The graphs of the current-phase dependences of a microcontact formed by superconducting magnesium diboride for different values of the ratio $r = R_{N1}/R_{N2}$ and temperature $\tau = T/T_c$ are shown in Fig. 20.

The temperature behavior of the critical current of the Josephson junction $I_c$ is shown in Fig. 21. It can be seen that,

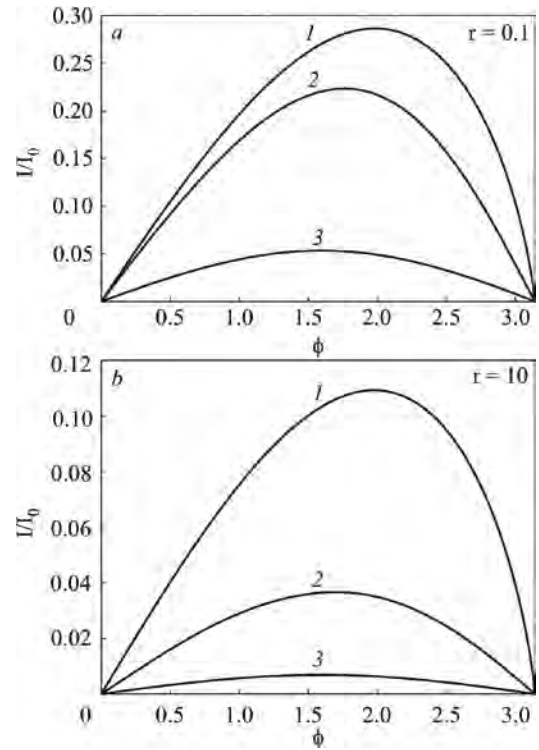

Fig. 20. Current-phase dependences of an SCS junction formed by MgB$_2$ on both sides, plotted at different temperatures $\tau = T/T_c$: $\tau = 0$ (1), $\tau = 0.5$ (2), $\tau = 0.9$ (3) and the ratios $r = R_{N1}R_{N2}$.[111]

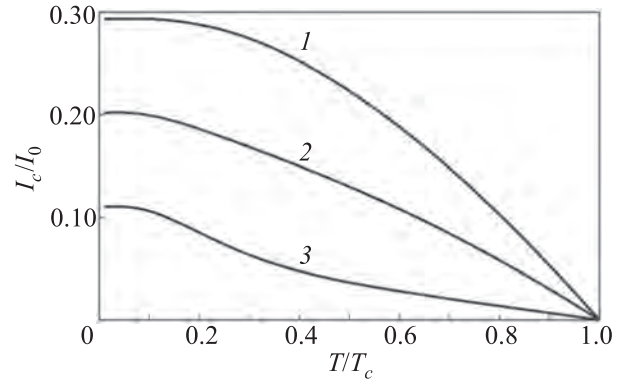

Fig. 21. Temperature dependence of the critical current of an ScS junction $I_c(T)$ for various ratios $r = R_{N1}/R_{N2}$.[111]

depending on the value of $r$, the curves $I_c(T)$ have both positive and negative curvature.

An interesting case is the intermixing of various band contributions, arising as a result of interband scattering. For arbitrary values and an arbitrary temperature $T$, the Usadel equations can be solved numerically. However, in order to first clarify the effect of interband scattering, the Josephson effect for an ScS contact can be considered near the critical temperature.[110] In this approximation, by solving the linearized Usadel equations, the current density through the system has the form

$$j = j_{11} + j_{22} + j_{12} + j_{21}, \quad (18)$$

where $j$ consists of four components arising due to transitions from the right bank to the left bank between different bands

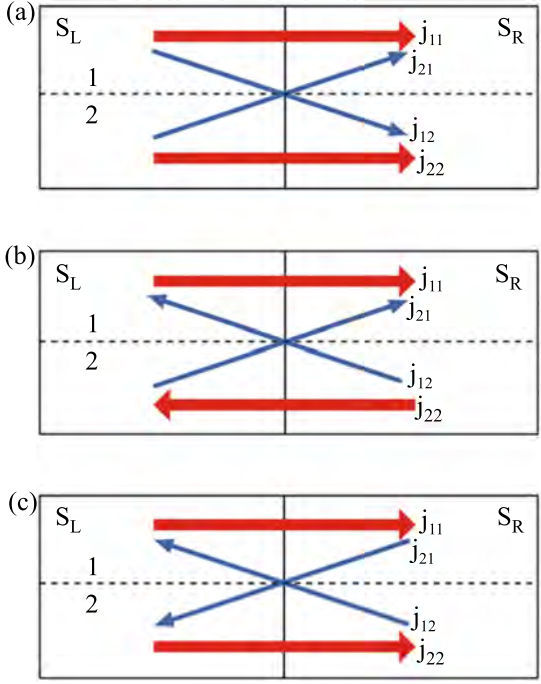

Fig. 22. Current flow directions in an ScS junction of two-band superconductors for different phase shifts in the banks: (a) phase difference of the order parameters in the left and right banks is 0; (b) there is a π-shift in the right superconducting bank, while the phase difference in the left banks is 0; and (c) the phase difference of the order parameters is π in both banks.[110]

$$j_{11} \sim |\Delta_1|^2 \sin\left(\phi_1^R - \phi_1^L\right),$$
$$j_{22} \sim |\Delta_2|^2 \sin\left(\phi_2^R - \phi_2^L\right),$$
$$j_{12} \sim |\Delta_1||\Delta_2| \sin\left(\phi_2^R - \phi_1^L\right),$$
$$j_{21} \sim |\Delta_1||\Delta_2| \sin\left(\phi_1^R - \phi_2^L\right).$$
(19)

The relative direction of the current density components $j_{ik}$ and the total current density depend on the internal phase difference of the order parameters in the banks $\delta\phi^{L,R}$ (Fig. 22). For $\delta\phi^L = 0$ and $\delta\phi^R = 0$ (both superconductors have the $s_{++}$-wave symmetry), the Josephson current is expressed as

$$I = \frac{\pi T}{eR_N(N_1 D_1 + N_2 D_2)}$$
$$\times \left( N_1 D_1 \sum_\omega^{\omega_D} \frac{(|\Delta_1|(\omega+\Gamma_{21}) + |\Delta_2|\Gamma_{12})^2}{(\omega^2 + (\Gamma_{12}+\Gamma_{21})\omega)^2} \right.$$
$$\left. + N_2 D_2 \sum_\omega^{\omega_D} \frac{(|\Delta_2|(\omega+\Gamma_{12}) + |\Delta_1|\Gamma_{21})^2}{(\omega^2 + (\Gamma_{12}+\Gamma_{21})\omega)^2} \right) \sin\phi = I_c \sin\phi,$$
(20)

where $\Gamma_{ij}$ corresponds to the interband scattering coefficients (microscopic analogue of the coefficient $\eta$).

For $\delta\phi^L = \pi$ and $\delta\phi^R = \pi$ (contacting superconductors with the $s_\pm$ symmetry), the current is equal to

$$I = \frac{\pi T}{eR_N(N_1 D_1 + N_2 D_2)}$$
$$\times \left( N_1 D_1 \sum_\omega^{\omega_D} \frac{(|\Delta_1|(\omega+\Gamma_{21}) - |\Delta_2|\Gamma_{12})^2}{(\omega^2 + (\Gamma_{12}+\Gamma_{21})\omega)^2} \right.$$
$$\left. + N_2 D_2 \sum_\omega^{\omega_D} \frac{(|\Delta_2|(\omega+\Gamma_{12}) - |\Delta_1|\Gamma_{21})^2}{(\omega^2 + (\Gamma_{12}+\Gamma_{21})\omega)^2} \right) \sin\phi = I_c \sin\phi.$$
(21)

Finally, the most interesting case, when the microcontact is formed by two-band superconductors with the $s_{++}$ and $s_\pm$ wave symmetries, i.e., $\delta\phi^L = 0$ and $\delta\phi^R = \pi$

$$I = \frac{\pi T}{eR_N(N_1 D_1 + N_2 D_2)}$$
$$\times \left( N_1 D_1 \sum_\omega^{\omega_D} \frac{\Delta_1^2(\omega+\Gamma_{21})^2 - \Delta_2^2 \Gamma_{12}^2}{(\omega^2 + (\Gamma_{12}+\Gamma_{21})\omega)^2} \right.$$
$$\left. + N_2 D_2 \sum_\omega^{\omega_D} \frac{\Delta_1^2 \Gamma_{21}^2 - \Delta_2^2(\omega+\Gamma_{12})^2}{(\omega^2 + (\Gamma_{12}+\Gamma_{21})\omega)^2} \right) \sin\phi = I_c \sin\phi.$$
(22)

It is obvious that for certain values of the constants $\Gamma_{12,21}$, the ratio $N_2 D_2/N_1 D_1$ and the gaps $|\Delta_1|$ and $|\Delta_2|$, Eq. (22) allows the existence of a negative critical current of the Josephson junction $I_c$. This means that the current-phase characteristic corresponds to the π-contact.

At an arbitrary temperature $0 \leq T \leq T_c$, the behavior of the ScS contact with the interband scattering taken into account can be studied using the perturbation theory in $\Gamma_{ij}$. In the first approximation, the anomalous Green's functions for each shore have the form:[111]

$$\begin{cases} f_1 = \dfrac{\Delta_1}{\sqrt{|\Delta_1|^2 + \omega^2}} + \Gamma_{12} \dfrac{(2\omega^2 + |\Delta_1|^2)(\Delta_2 - \Delta_1) - \Delta_1^2(\Delta_2^* - \Delta_1^*)}{2\sqrt{(|\Delta_1|^2 + \omega^2)^3}\sqrt{|\Delta_2|^2 + \omega^2}}, \\ f_2 = \dfrac{\Delta_2}{\sqrt{|\Delta_2|^2 + \omega^2}} + \Gamma_{21} \dfrac{(2\omega^2 + |\Delta_1|^2)(\Delta_1 - \Delta_2) - \Delta_2^2(\Delta_1^* - \Delta_2^*)}{2\sqrt{|\Delta_1|^2 + \omega^2}\sqrt{(|\Delta_2|^2 + \omega^2)^3}}. \end{cases}$$
(23)

It follows from Eq. (23) that the presence of interband impurities leads to the fact that the phases of the anomalous Green's functions $f_i$ do not coincide with the phases of the order parameters $\Delta_i$. Correspondingly, the corrections to the current due to weak interband scattering have the form

$$\delta I = \delta I_1 + \delta I_2 \tag{24}$$

$$\delta I_1 = \frac{2\pi T \Gamma_{21}}{eR_{N1}}$$
$$\times \sum_\omega \left( \frac{\omega^2(|\Delta_2|e^{i\delta} - |\Delta_1|)\cos\left(\dfrac{\phi}{2}\right)}{\sqrt{\left(|\Delta_1|^2 \cos^2\left(\dfrac{\phi}{2}\right) + \omega^2\right)^3}\sqrt{|\Delta_2|^2 + \omega^2}} \right.$$
$$\times \arctan \frac{|\Delta_1|\sin\dfrac{\phi}{2}}{\sqrt{\omega^2 + |\Delta_1|^2 \cos^2\dfrac{\phi}{2}}}$$
$$\left. + \frac{1}{2}\frac{\omega^2 |\Delta_1|(|\Delta_2|e^{i\delta} - |\Delta_1|)\sin\phi}{(|\Delta_1|^2 + \omega^2)\left(|\Delta_1|^2 \cos^2\left(\dfrac{\phi}{2}\right) + \omega^2\right)\sqrt{|\Delta_2|^2 + \omega^2}} \right),$$
(25)

$$\delta I_2 = \frac{2\pi T \Gamma_{21}}{eR_{N2}}$$

$$\times \sum_\omega \left( \frac{\omega^2 (|\Delta_1| - e^{i\delta}|\Delta_2|)\cos\left(\frac{\phi}{2}\right)}{\sqrt{\left(|\Delta_2|^2\cos^2\left(\frac{\phi}{2}\right) + \omega^2\right)^3}\sqrt{|\Delta_1|^2 + \omega^2}} \right.$$

$$\times \arctan\frac{|\Delta_2|\sin\frac{\phi}{2}}{\sqrt{\omega^2 + |\Delta_2|^2\cos^2\frac{\phi}{2}}}$$

$$\left. + \frac{1}{2}\frac{\omega^2|\Delta_2|(|\Delta_1| - e^{i\delta}|\Delta_2|)\sin\phi}{(|\Delta_2|^2 + \omega^2)\left(|\Delta_2|^2\cos^2\left(\frac{\phi}{2}\right) + \omega^2\right)\sqrt{|\Delta_1|^2 + \omega^2}} \right). \quad (26)$$

The magnitude of the corrections depends on the symmetry type of the superconducting banks $\delta = 0$ or $\pi$, and the structure of the obtained expressions unambiguously indicates the entanglement of currents from different zones. The same result was obtained earlier in the study of the characteristics of a microchip near the critical temperature[110] (see Fig. 22).

The behavior of the Josephson ScS system becomes substantially more complicated when single-band and three-band superconductors are in contact. This is due to the fact that in a three-band superconductor, even when all three order parameters have s-wave symmetry, for certain magnitudes of the interband interaction, a frustration phenomenon occurs, leading to degeneracy of the ground state and, as a consequence, to time reversal symmetry breaking. Earlier, it has been predicted that this state can appear as a result of the anomalous proximity effect between $s_{++}$ and $s_\pm$ superconducting layers, as well as in the Josephson junction formed by single-band and two-band superconductors with $s_{++}$ and $s_\pm$ wave symmetry of the order parameter, respectively.[63,64,66,67] However, in this case, this phenomenon initially takes place in the massive banks of the Josephson system.

In Ref. 112, the effect of time reversal symmetry breaking on the Josephson effect has been studied in a three-band superconductor forming an ScS contact together with a single-band superconducting bank (Fig. 23). The analysis has been carried out in the ballistic limit using the formalism of the Bogolyubov-de Gennes equations, the solution of which in the simplest case when the energy gaps of single-band and three-band superconductors coincide gives an expression for the Josephson current:

$$I(\varphi) = \frac{e|\Delta|}{\hbar}\left\{\sin\frac{\varphi}{2}\tanh\frac{|\Delta|\cos\frac{\varphi}{2}}{2T} + \sin\frac{\varphi + \varphi_{21}}{2}\right.$$

$$\times \tanh\frac{|\Delta|\cos\frac{\varphi + \varphi_{21}}{2}}{2T} + \sin\frac{\varphi + \varphi_{31}}{2}$$

$$\left. \times \tanh\frac{|\Delta|\cos\frac{\varphi + \varphi_{31}}{2}}{2T} \right\}, \quad (27)$$

where $|\Delta|$ is the energy gap, $\varphi$ corresponds to the Josephson phase difference (the phase difference between the order parameter phase of the single-band superconductor and the phase of the first order parameter in the three-band superconductor), and $\varphi_{21}$ and $\varphi_{31}$ are the interband phase differences in the three-band superconductor, which determine the position of its ground state.

As can be seen from the form of Eq. (27), it is actually a generalization of the KO-2 theory to the case of three gaps.

Figure 24 shows the current-phase dependence of a microcontact in the case when the three-band superconductor is in a state with broken time reversal symmetry. Based on the shape of the current-phase dependence, it is possible to draw the main conclusion about the features of an ScS contact between a single-band and three-band superconductors: the critical currents of the system are not equivalent for two opposite directions. The reason is that the current-phase dependence does not satisfy the oddness property, $I(-\varphi) \neq -I(\varphi)$, which is caused by the time reversal symmetry breaking. The asymmetry of the critical currents can also be established using the absence of symmetry in the Andreev spectra (see the inset in Fig. 24).

With increasing temperature, the translational antisymmetry, which is inherent in ordinary Josephson junctions, is restored, and the critical currents of different directions become equal to each other. In other words, the main difference of the ScS microcontacts involving a three-band superconductor with time-reversal symmetry breaking is most pronounced at low temperatures.

In Ref. 112, the features of the ac Josephson effect, in particular the microwave response of a contact formed by single- and three-band superconducting bands, have been also analyzed. For the illustration purposes, the case of coincident gaps has again been considered, while the ground state of the three-band superconductor was fixed in the form $\varphi_{21} = 2\pi/3$ and $\phi_{31} = -2\pi/3$. Calculations showed that the current-voltage characteristic of the microcontact contains fractional Shapiro steps of the, which are highest at the frequencies multiple of 1/3 of the applied voltage frequency (Fig. 25).

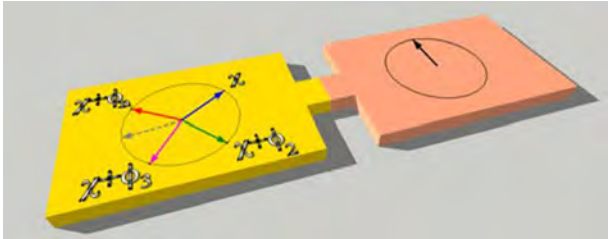

Fig. 23. Schematic representation of an ScS microcontact between a single-band and multi-band (three-band) superconductor with time-reversal symmetry breaking. The arrows correspond to the phases of the order parameters in each of the superconducting banks.[114]

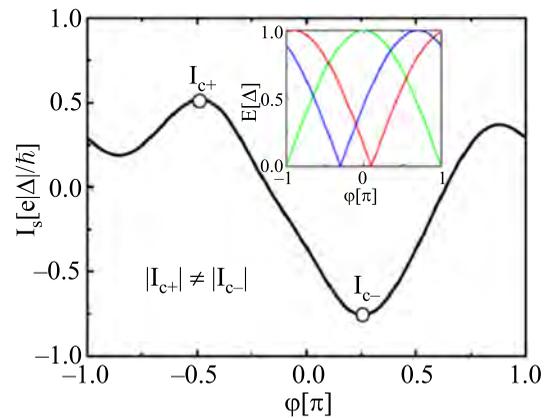

Fig. 24. Current-phase dependence of a ballistic microcontact formed by a single-band superconductor and a three-band superconductor with time-reversal symmetry breaking and the ground state parameters $\varphi_{21} = 0.9\pi$ and $\varphi_{31} = 1.3\pi$ and temperature $T = 0.2|\Delta|$. The inset shows the Andreev spectrum of the microcontact for the same parameters as the current-phase dependence.[112]

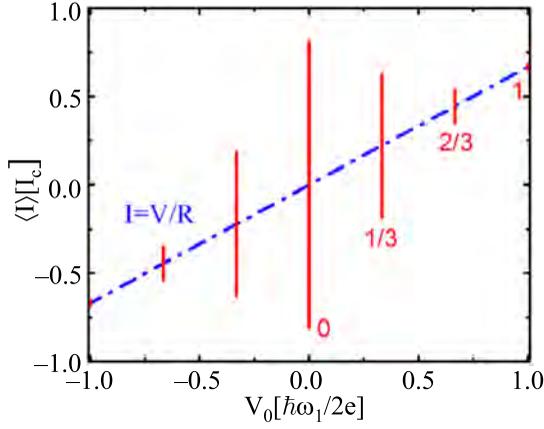

Fig. 25. Schematic view of the current-voltage characteristic with fractional Shapiro steps in a microcontact formed by a single-band superconductor and a three-band superconductor with time-reversal symmetry breaking and the ground state parameters $\varphi_{21} = 2\pi/3$ and $\varphi_{31} = -2\pi/3$. The microcontact is controlled by an ac voltage of frequency $\omega_1$.[112]

The opposite case, namely, the diffusion limit of a microcontact formed by a single-band and three-band superconductor with time-reversal symmetry breaking, has been considered in Ref. 113. As in the KO-1 theory,[102] the point-contact geometry allows to reduce the problem to one-dimensional Usadel equations, generalized only to three energy gaps, in which all terms but the gradient are discarded. Since the analytic expression for the Josephson current generally looks rather cumbersome, in this paper it was assumed for simplicity that the energy gaps in the contacting superconductors coincide, and the study of the characteristics is carried out only for two temperature regimes: $T = 0$ and $T \to T_c$ (critical the temperatures of the superconductors are approximately equal). In the framework of these approximations, the current is represented in the form

$$I = \frac{\pi|\Delta|}{eR_{N1}} \cos\frac{\chi}{2} \operatorname{arctanh} \sin\frac{\chi}{2}$$
$$+ \frac{\pi|\Delta|}{eR_{N2}} \cos\frac{\chi+\phi}{2} \operatorname{arctanh} \sin\frac{\chi+\phi}{2}$$
$$+ \frac{\pi|\Delta|}{eR_{N3}} \cos\frac{\chi+\theta}{2} \operatorname{arctanh} \sin\frac{\chi+\theta}{2}, \quad (28)$$

and the energy is

$$E = \frac{|\Delta|\Phi_0}{2eR_{N1}}\left(2\sin\frac{\chi}{2}\operatorname{arctanh}\sin\frac{\chi}{2} + \ln\cos^2\frac{\chi}{2}\right)$$
$$+ \frac{|\Delta|\Phi_0}{2eR_{N2}}\left(2\sin\frac{\chi+\phi}{2}\operatorname{arctanh}\sin\frac{\chi+\phi}{2} + \ln\cos^2\frac{\chi+\phi}{2}\right)$$
$$+ \frac{|\Delta|\Phi_0}{2eR_{N3}}\left(2\sin\frac{\chi+\theta}{2}\operatorname{arctanh}\sin\frac{\chi+\theta}{2} + \ln\cos^2\frac{\chi+\theta}{2}\right), \quad (29)$$

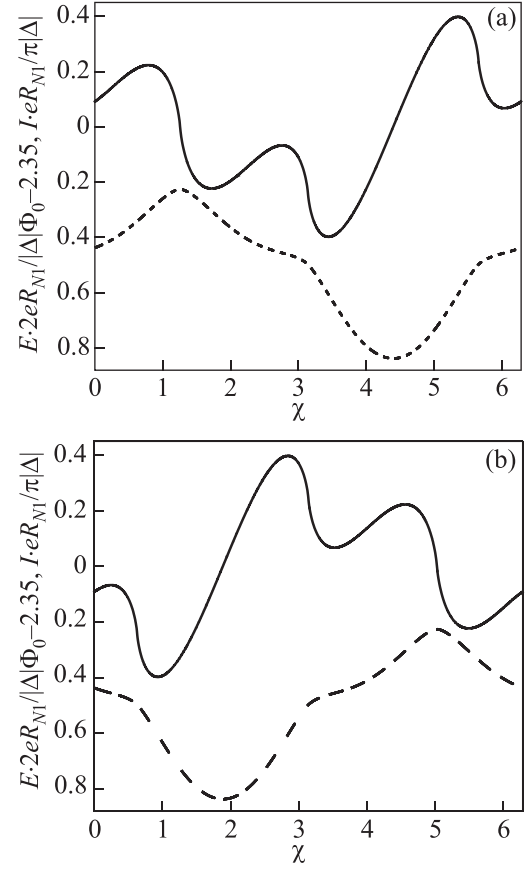

Fig. 26. Current-phase dependences (solid line) and the Josephson energy (dashed line) of microcontacts at zero temperature between a single-band superconductor and three-band superconductor with time-reversal symmetry breaking with the phase difference $\phi = 0.6\pi$, $\theta = 1.2\pi$ (a) and $\phi = 1.4\pi$, $\theta = 0.8\pi$ (b). The ratio of the partial contributions of each band to the normal contact resistance $R_{N1}/R_{N2} = R_{N1}/R_{N3} = 1$. The energy gaps of the contacting superconducting banks coincide.[113]

where $\chi$ determines the Josephson phase difference and $\phi$ and $\theta$ determine the ground state (the phase difference of the order parameters) of the three-band superconductor.

The time-reversal symmetry breaking in a three-band superconductor creates complex asymmetric current-phase dependences of the microcontact (Fig. 26), and the position of the energy minimum of the Josephson system explicitly indicates the realization of the $\varphi$-contact. At the same time, the realization of a specific type of the current-phase dependence which is observed in the course of the experimental measurements depends on the "prehistory" of the three-band superconducting bank, i.e., in which of the two basic states it "falls."

In the temperature limit close to critical, the additive contributions to the Josephson current acquire a sinusoidal dependence, but despite this, the ScS system retains the singularity in the form of a $\varphi$-contact. A further analysis of the microcontact behavior has shown that if the time-reversal symmetry is violated in the three-band superconductor, then the current-phase characteristic has the $\varphi$-contact properties in the entire temperature interval. For a three-band superconductor in which the ground state does not become frustrated, the $\varphi$-contact observed at $T = 0$ evolves as the temperature increases either into ordinary or $\pi$-contact.

The KO-2 theory, known as a model of a microcontact with ballistic conductivity between single-band superconductors, was generalized in Ref. 114 to the case of the Josephson system formed by single-band and three-band superconducting shores. At a qualitative level, this behavior coincides with the behavior observed for a diffusion microcontact. The differences are found only in the phase diagram showing the total number of energy minima for a given system as a function of the phase difference of the order parameters in a massive three-band superconductor.

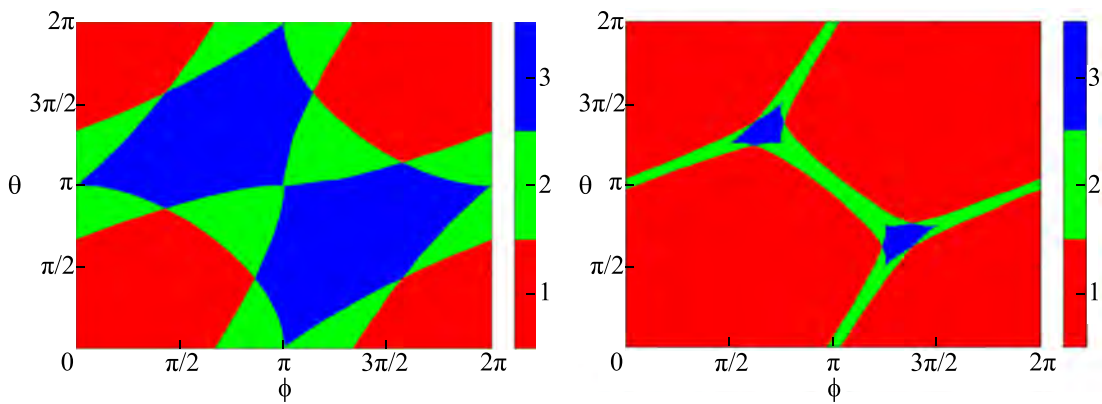

Fig. 27. Dependence of the total number of global and local energy minima of the Josephson junction with ballistic (left) and diffusion (right) conductivity between a single-band and three-band superconductors depending on the "position" of the ground state (values of $\phi$ and $\theta$).[114]

The main feature of a system with ballistic conductivity between a single-band and three-band superconductors is a substantially wider variety of the states of the Josephson junction (Fig. 27, left), in comparison with a similar system in the dirty limit (Fig. 27, right). In other words, for a diffusion microcontact, the existence intervals for additional (local) minima are much smaller.

In a microcontact formed by identical three-band superconductors with time reversal symmetry broken, the critical current can be significantly suppressed when superconducting banks are in different ground states compared to the case of identical ground states.[115] From the experimental point of view, when the superconducting state "reset" and the cooling process repeated, measurements of the critical current can produce different values.

### 3.3. Quantum interferometers based on multiband superconductors

One of the main motivations for studying the Josephson effect as well as the proximity effect in multiband superconductors, is to obtain information about the symmetry of the order parameter, in the context of a new iron-based family of superconducting compounds. In the previous sections it was shown how, using the assumption of the unusual form of the wave function of Cooper pairs, one can theoretically predict the features in the behavior of different Josephson junctions based on superconductors with multiple energy gaps. The detection of these distinctive features is the basis for a technique known as the Josephson interferometry (see, for example, Ref. 116).

The Josephson interferometry includes the study of the magnetic response of a contact, its current-phase dependences, as well as SQUID interferometry. Among this variety of methods, the last one is often the most useful. From a technical point of view, it is based on the study of the characteristics of a single-contact interferometer (a Josephson junction in a superconducting ring) or a dc SQUID (a superconducting ring with two Josephson junctions). In this geometry, one of the superconductors in contact usually has isotropic s-wave symmetry of the order parameter, and the second is a multi-band superconductor, the symmetry of the order parameter in which is under study.[117,118]

To explain the results of the experiment on the magnetic response of a closed loop formed by a niobium filament and massive iron oxypnictide $NdFeAsO_{0.88}F_{0.12}$, in particular, the observations of transitions corresponding to half of the magnetic flux quantum,[38] in Ref. 67, the Ginzburg-Landau formalism with functional (4) has been applied to describe the properties of a composite $s$–$s_{\pm}$ ring (formally, a dc SQUID). Depending on the phenomenological parameters, the free Ginzburg-Landau energy as a function of the applied external magnetic flux exhibits three regimes: a regime with one minimum at $\Phi = n\Phi_0$ (blue curve); a regime with broken time-reversal symmetry, which has a degenerate ground state at $\Phi = n\Phi_0 \pm \Delta\Phi$, where the shift $\Delta\Phi$ is determined by the parameters of the s–$s_{\pm}$ SQUID (green curve); and the regime with two minima, one of which is a local (red curve), realized at $\Phi = \left(n + \frac{1}{2}\right)\Phi_0$ (Fig. 28).

Proceeding from this theoretical model, the above-mentioned experiment can be explained as follows: as the critical current in the ring increases, the probability of a system to jump from the regime with one minimum at $\Phi = n\Phi_0$ to the regime with two minima and metastable state at $\Phi = \left(n + \tfrac{1}{2}\right)\Phi_0$ increases. The probability of staying in this local minimum is less than in the ground state with an integer flux quantum. Therefore, if the system receives a push and goes into a state with a half-integer flux quantum, then, most likely, it will be end up in a state with $\Phi = n\Phi_0$. This means that quantum jumps with half-integer flux quantum correlate and, as a rule, appear in pairs, which is observed experimentally.

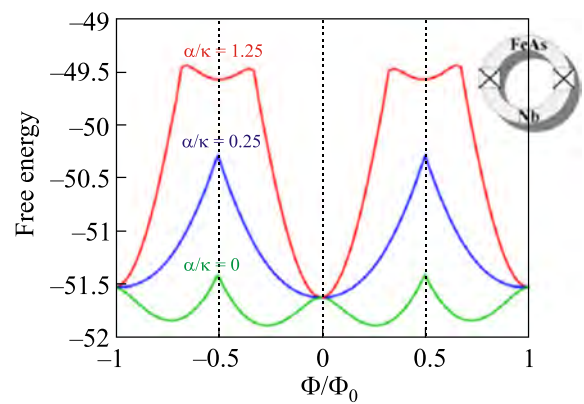

Fig. 28. Energy of dc SQUID formed by s and $s_{\pm}$-wave superconducting halves as a function of the applied external magnetic flux for different parameters of the two-band superconductor, described by functional (4).[67]

A similar behavior in a dc SQUID based on s and $s_{\pm}$-wave superconductors has also been confirmed in the framework of a self-consistent microscopic description using the Bogolyubov-de Gennes equation.[68] In addition to the three regimes that were found earlier using the phenomenological Ginzburg-Landau approach, the paper has also revealed the existence of another state, when a single global minimum is realized at $\Phi = \left(n + \tfrac{1}{2}\right)\Phi_0$.

As in the case of the Josephson effect, a dc SQUID formed by s-wave single-band and three-band superconductors, gives rise to a number of fundamentally new features of macroscopic quantum interference that are not observed in other similar systems based on two-band and other superconducting compounds with an unusual type of symmetry of the order parameter. This has first been shown theoretically in Ref. 119, in which a dc SQUID with diffusion microcontacts between single-band and three-band superconductors has been considered.

Figure 29 shows the contour maps of the energy as a function of the phase differences $\chi_1$ and $\chi_2$ at the junctions for an ordinary dc SQUID formed by s-wave superconductors [Figs. 29(a) and 29(b)] and a dc SQUID based on a three-band superconductor with broken time-reversal symmetry [Figs. 29(c)–29(e)] and without symmetry breaking [Figs. 29(g)–29(i)] for zero magnetic flux (left column) and for half-integer flux (right column) in the zero-temperature limit.

In the case of a three-band superconductor with time-reversal symmetry breaking at zero magnetic flux, only a shift in the minimum position of the dc SQUID from the zero point $\chi_1, \chi_2 = 0$ is observed [as is the case for an ordinary dc SQUID in Fig. 29(a)], in spite of the presence of a frustrated ground state [Figs. 29(c) and 29(e)]. The main difference between a dc SQUID based on a three-band superconductor with broken time-reversal symmetry [Figs. 29(d) and 29(f)] and an ordinary dc SQUID [Fig. 29(b)] is strong degeneracy of the ground state at half-integer magnetic flux quanta.

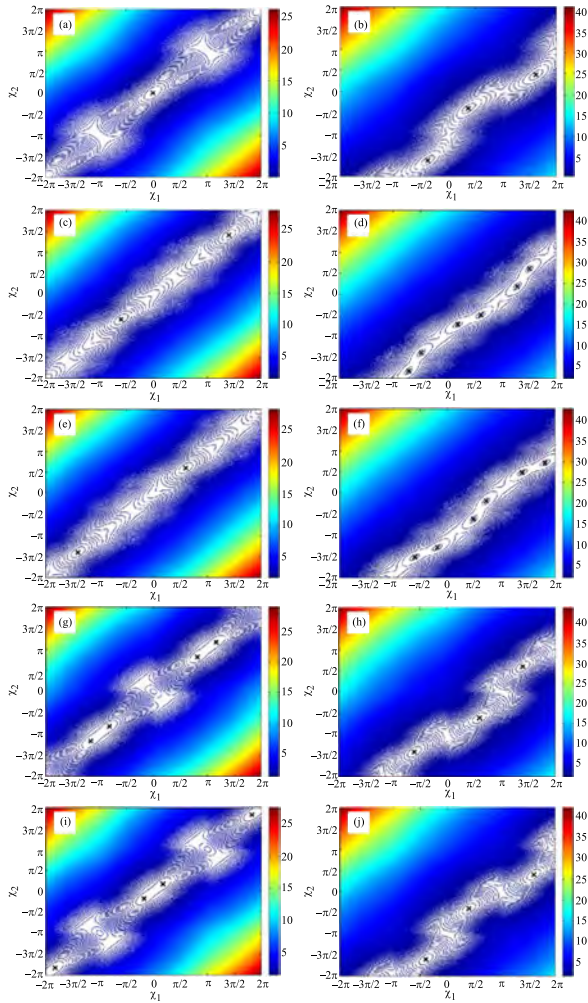

Fig. 29. Contour map of the energy surface of a dc SQUID for zero external magnetic flux $\chi_e = 0$ ($\Phi/\Phi_0 = 0$, left column) and for $\chi_e = \pi$ ($\Phi/\Phi_0 = 0.5$, right column) in the absence transport current. Panels (a) and (b) are constructed for microcontacts between s-wave single-band superconductors; (c), (d) and (e), (f) correspond to the microcontacts between a single-band superconductor and a three-band superconductor with time-reversal symmetry breaking for the frustrated ground states $\varphi = 0.6\pi$, $\theta = 1.2\pi$ and $\varphi = 1.4\pi$, $\theta = 0.8\pi$; (g), (h) and (i), (j) correspond to the microcontacts between a single-band and three-band superconductors without time-reversal symmetry breaking with the ground state $\varphi = \pi$, $\theta = \pi$ and $\varphi = 0$, $\theta = \pi$. Crosses indicate the positions of global minima.[119]

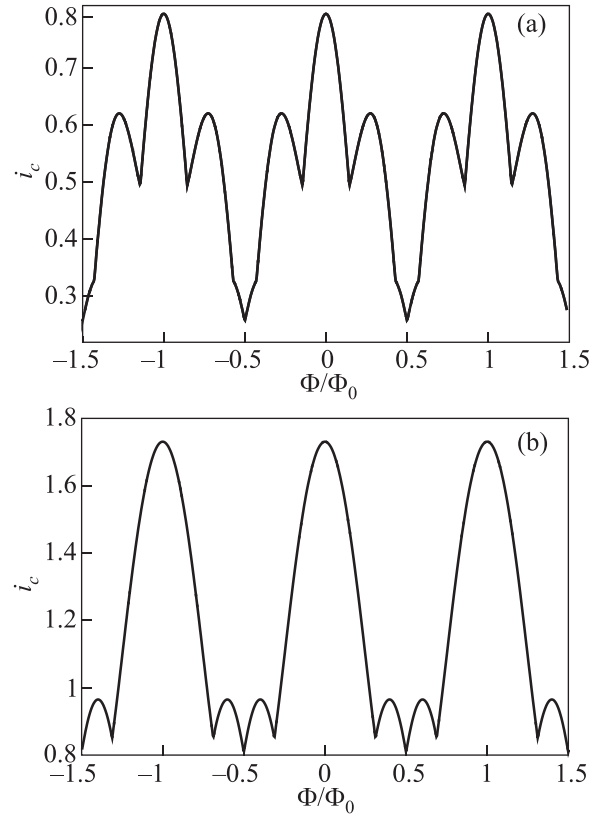

Fig. 30. Dependence of the critical current of a symmetric dc SQUID on the applied magnetic flux for a three-band superconductor with time-reversal symmetry breaking with the frustrated ground states $\phi = 0.6\pi$, $\theta = 1.2\pi$ and $\phi = 1.4\pi$, $\theta = 0.8\pi$ (a) and a three-band superconductor without time-reversal symmetry breaking with the ground state $\phi = 0$, $\theta = \pi$ and $\phi = \pi$, $\theta = \pi$ (b).[119]

For a dc SQUID with a three-band superconductor in which the time-reversal symmetry is not broken the opposite behavior is realized. At zero magnetic flux, the ground state is degenerate [Figs. 29(g) and 29(i)]. However, at a magnetic flux corresponding to half-integer flux quanta, this degeneracy is lifted [Figs. 29(h) and 29(j)], and the system behaves as an ordinary dc SQUID.

One of the most important characteristics of SQUID is the dependence of the critical current $i_c$ on the external magnetic flux $\Phi_e$. A graph of these dependences for a symmetric system (Josephson junctions have identical critical currents) is shown in Fig. 30.

Despite the presence of two possible current-phase relations in Josephson junctions,[113] the $i_c = i_c(\Phi_e/\Phi_0)$ dependence is the same for both ground states [Fig. 30(a)]. The same situation is realized for a three-band superconductor without time-reversal symmetry breaking with ground states at the phases $\varphi = 0, \theta = \pi$ and $\varphi = \pi, \theta = \pi$ [Fig. 30(b)]. As follows from Fig. 30, the critical current for a three-band superconductor with time-reversal symmetry breaking has more pronounced lateral peaks on the $i_c = i_c(\Phi_e/\Phi_0)$ dependence.

The introduction of an asymmetry of the critical currents of the Josephson microcontacts in a dc SQUID leads to an expected asymmetry of the $i_c = i_c(\Phi_e/\Phi_0)$ dependences (Fig. 31). This effect is especially noticeable for a three-band superconductor with time-reversal symmetry breaking [Fig. 31(a)] and without it [Fig. 31(b)].

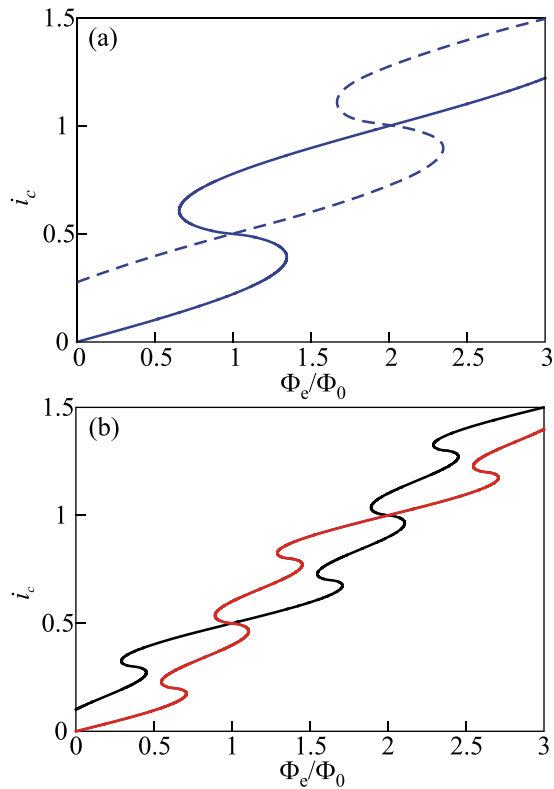

Fig. 32. S-states in an ordinary dc SQUID (a) and a dc SQUID based on a three-band superconductor with time-reversal symmetry breaking (b). The blue solid and dashed curves show two possible types of S-states in an ordinary dc SQUID. The dependences $\Phi(\Phi_e)$ for the case of a three-band superconductor with time-reversal symmetry breaking correspond to the frustrated ground state $\phi = 0.6\pi, \theta = 1.2\pi$ (black curve) and $\phi = 1.4\pi, \theta = 0.8\pi$ (red curve).[119]

Furthermore, in Ref. 119, the S-states of a dc SQUID have also been studied: the dependence of the total magnetic flux through the loop on the external magnetic flux at zero transport current. In comparison with the hysteretic behavior of a conventional dc SQUID, a similar system based on a three-band superconductor with time-reversal symmetry breaking exhibits a multihysteresis regime (Fig. 32).

In other words, in the process of measuring the total magnetic flux as a function of the applied flux, additional jumps in these dependences should occur. The latter can be regarded as the main feature of a dc SQUID with diffusion microcontacts between a single-band and three-band superconductors with time-reversal symmetry breaking.

For a dc SQUID formed by the Josephson microcontacts between a single-band and multi-band (two- or three-band) superconductors with ballistic conductivity, the specific features of the system are preserved, differing only quantitatively.[114]

## 4. Experimental results

Despite the predominantly theoretical nature of this review, it would be unfair not to mention experimental work in this direction. Josephson structures based on magnesium diboride have been studied in detail in Refs. 120–155. The two-band superconductor model with $s_{++}$ wave symmetry,

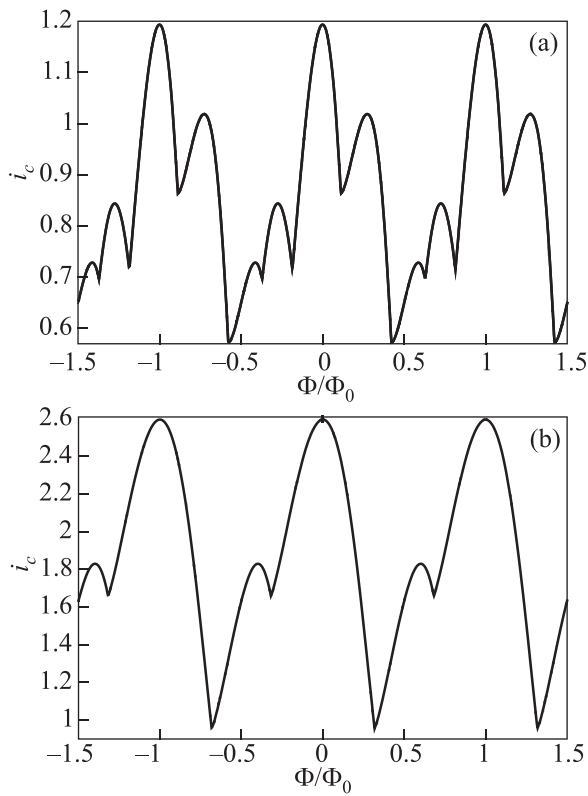

Fig. 31. Dependence of the critical current of an asymmetric dc SQUID on the applied external magnetic flux for a three-band superconductor with time-reversal symmetry breaking with frustrated ground states $\phi = 0.6\pi, \theta = 1.2\pi$ and $\phi = 1.4\pi, \theta = 0.8\pi$ (a) and a three-band superconductor without time-reversal symmetry breaking with the ground state $\phi = 0, \theta = \pi$ and $\phi = \pi, \theta = \pi$ (b).[119]

which is realized in MgB$_2$, describes most of the experimental data well.

In iron-containing superconductors, where probably s$_\pm$ or more exotic chiral symmetry of the order parameter is realized, the Josephson effect has first been observed in Ref. 156. The objects under study were planar and microcontacts formed by ordinary superconducting lead and Ba$_{1-x}$K$_x$Fe$_2$As$_2$ single crystal. The measured I-V characteristics of the junction showed no significant hysteresis and were well described by the RSJ model.

Under microwave irradiation, Shapiro steps were also observed on the I-V characteristics, which appeared at voltages corresponding to a multiple frequency of the external stimulus. In addition, there was a symmetric Fraunhofer picture of the modulation of the junction critical current in an external magnetic field. From the experimental facts obtained, a preliminary conclusion has been drawn that the order parameter in Ba$_{1-x}$K$_x$Fe$_2$As$_2$ at least does not demonstrate the presence of zeros (nodes) and has s-wave symmetry.

The experimental data on the proximity effect and Josephson effects in oxypnictides and iron chalcogenides performed before 2011 were summarized in review Ref. 157. Subsequent experiments have confirmed the multi-band structure of "iron" superconductors,[158–165] while some of the results,[163] specifically, the temperature behavior of the junction critical current, can be interpreted as the theoretically predicted possibility of a 0-$\pi$ transition. More exotic predictions described in the review have not, to our knowledge, been confirmed experimentally so far.

## 5. Conclusion

Thus, in this review we considered the proximity effect and the Josephson effects in various heterostructures in which one of the components was a multi-band superconductor. It was established that in the bilayer of s and s$_\pm$ superconductors, peculiar features can appear in the form of a state with broken time-reversal symmetry, which manifest themselves in the form of unique peaks and dips in the density of states at the bilayer interface. It was shown that the magnitude of these peaks, as well as the dips, is governed by the phase difference of the order parameters in the two-band superconductor, induced by time-reversal symmetry breaking.

The SIS, SNS and ScS Josephson junctions exhibit a substantially richer behavior. For systems in which one or both superconducting banks consist of an s$_\pm$ two-band or three-band superconductor, depending on the temperature as well as the type and parameters of the barrier, it is possible to realize an ordinary junction with the ground state with a Josephson phase difference equal to zero, a $\pi$-junction with the ground state that has a phase difference of $\pi$ and an exotic $\varphi$-contact, in which the ground state can become frustrated, leading to time-reversal symmetry breaking. It was shown that in the case of a $\pi$-contact or its "generalization" in the form of a $\varphi$-contact, the temperature dependence of the critical current exhibits kinks, and the current-phase dependence can acquire a complex shape, showing the presence of spontaneous currents. Moreover, it was found that if one of the superconducting banks of the junction is a three-band superconductor with time-reversal symmetry breaking, two types of current-phase dependences can be possible observed in an experiment. The experimentally observed dependence is determined by which of the two ground states the bulk three-band superconductor was in prior the measurement. If the Josephson junction is created by identical three-band superconductors with broken time-reversal symmetry, the critical current of the system can vary its magnitude from the maximum corresponding to the same ground state of both superconducting banks to the minimum occurring when the ground states are complex conjugate.

The multi-band nature of the superconducting state also affects the ac Josephson effect. It was shown that the fractional Shapiro steps appear on the I-V characteristic under microwave irradiation of the junction, and the symmetry of the order parameter (s$_{++}$, s$_\pm$ or chirality in the case of a three-band superconductor) governs the unique structure in the arrangement of these steps.

It was found that the above features of the Josephson effect are also extended to the systems of Josephson junctions, in particular dc SQUIDs formed by a conventional and multi-band superconductors. The presence of s$_\pm$ wave symmetry in a two-band superconductor or time-reversal symmetry breaking in a three-band superconductor forms a unique energy spectrum of the SQUID, which is reflected in the unusual dependences of the critical current on the applied external magnetic flux and the specific shape of the S-states.

This work was supported by Russian Science Foundation, Grant No. 15-12-10020.

## APPENDIX: MICROSCOPIC AND PHENOMENOLOGICAL THEORY OF MULTIBAND SUPERCONDUCTIVITY

For the theoretical investigation of current states in multi-band superconductors, the quasiclassical formalism of the Eilenberger equations and their approximation in the case of strong scattering by impurities in the form of Usadel equations with a corresponding generalization into several bands are used in most problems.

From these equations, in the limit of temperatures close to critical, the Ginzburg-Landau functional and equations for a superconductor with a multicomponent order parameter can be obtained by standard procedures. In this Appendix, we present a microscopic theory of multiband superconductivity based on the Eilenberger and Usadel equations, as well as the phenomenological formalism of the Ginzburg-Landau theory under the assumption that the wave function of Cooper pairs is isotropic (the order parameters have s-wave symmetry).

### A. Eilenberger and Usadel equations

The Eilenberger equations describing an anisotropic two-band superconductor for the quasiclassical Green functions $f_i(\mathbf{v}, \mathbf{r}, \omega)$, $f_i^*(\mathbf{v}, \mathbf{r}, \omega)$ and $g_i(\mathbf{v}, \mathbf{r}, \omega)$ in the general case have the form[166]

$$(2\omega + v_\mathbf{k}\mathbf{\Pi})f_1(\mathbf{k}) - 2\Delta_1 g_1(\mathbf{k}) = n_{\text{imp}} \int |u_{\mathbf{k}_1\mathbf{q}_1}|^2 [g_1(\mathbf{k}_1)f_1(\mathbf{q}_1) - f_1(\mathbf{k}_1)g_1(\mathbf{q}_1)] \frac{dA_{\mathbf{q}_1}}{v_{\mathbf{q}_1}}$$
$$+ n_{\text{imp}} \int |u_{\mathbf{k}_1\mathbf{q}_2}|^2 [g_1(\mathbf{k}_1)f_2(\mathbf{q}_2) - f_1(\mathbf{k}_1)g_2(\mathbf{q}_2)] \frac{dA_{\mathbf{q}_2}}{v_{\mathbf{q}_2}}, \qquad (A.1)$$

$$(2\omega + v_\mathbf{k}\mathbf{\Pi})f_2(\mathbf{k}) - 2\Delta_2 g_2(\mathbf{k}) = n_{\text{imp}} \int |u_{\mathbf{k}_2\mathbf{q}_2}|^2 [g_2(\mathbf{k}_2)f_2(\mathbf{q}_2) - f_2(\mathbf{k}_2)g_2(\mathbf{q}_2)] \frac{dA_{\mathbf{q}_1}}{v_{\mathbf{q}_1}}$$
$$+ n_{\text{imp}} \int |u_{\mathbf{k}_2\mathbf{q}_1}|^2 [g_2(\mathbf{k}_2)f_1(\mathbf{q}_1) - f_2(\mathbf{k}_2)g_1(\mathbf{q}_1)] \frac{dA_{\mathbf{q}_1}}{v_{\mathbf{q}_1}}, \qquad (A.2)$$

together with two more complex conjugate equations. Here $\mathbf{\Pi} \equiv \nabla + \frac{2\pi i}{\Phi_0}\mathbf{A}$, $\mathbf{A}$ is the vector potential of the magnetic field, $\Delta_i$ are the energy gaps, $\frac{dA_\mathbf{q}}{v_\mathbf{q}}$ denotes integration over the Fermi surface with a local density of states $\frac{1}{v_\mathbf{q}}$, the integrand terms take into account the scattering by nonmagnetic impurities, $u_{\mathbf{k}\mathbf{q}}$ is the scattering amplitude, $n_{\text{imp}}$ is the impurity density, $v_\mathbf{q}$ is the normal component of the group velocity on the anisotropic Fermi surface, the wave vectors $\mathbf{k}$ and $\mathbf{q}$ lie on the Fermi surface, $\omega = (2n+1)\pi T$ is the Matsubara frequency ($n \in \mathbb{Z}$) and the asterisk denotes complex conjugation. Equations (A.1) and (A.2) are supplemented by the normalization condition

$$ff^* + g^2 = 1 \qquad (A.3)$$

self-consistency equation for the superconducting gap

$$\Delta(\mathbf{k}, \mathbf{r}) = 2\pi T \sum_{\omega > 0}^{\omega_D} \int V(\mathbf{k}, \mathbf{q}) f(\mathbf{q}, \mathbf{r}, \omega) \frac{dA_\mathbf{q}}{v_\mathbf{q}}, \qquad (A.4)$$

and an expression for the superconducting current density

$$j = -4\pi Te \, \text{Im} \sum_{\omega > 0}^{\omega_D} \int v_\mathbf{q} g(\mathbf{q}, \mathbf{r}, \omega) \frac{dA_\mathbf{q}}{v_\mathbf{q}}, \qquad (A.5)$$

where $V(\mathbf{k}, \mathbf{q})$ is the pairing potential.

In the dirty limit (strong scattering on impurities), the Eilenberger equations are reduced to the Usadel equations for two-band superconductivity by expanding the Green's functions in spherical harmonics:

$$2\omega f_1 - D_1[g_1\Pi^2 f_1 - f_1\nabla^2 g_1] = 2\Delta_1 g_1 + \Gamma_{12}(g_1 f_2 - g_2 f_1), \qquad (A.6)$$

$$2\omega f_2 - D_2[g_2\Pi^2 f_2 - f_2\nabla^2 g_2] = 2\Delta_2 g_2 + \Gamma_{21}(g_2 f_1 - g_1 f_2), \qquad (A.7)$$

with the self-consistency equation

$$\Delta_i = 2\pi T \sum_j \sum_{\omega > 0}^{\omega_D} \lambda_{ij} f_j, \qquad (A.8)$$

which for the convenience of numerical calculation can be rewritten as

$$\begin{cases} 2\pi T \sum_{\omega>0} \left( \frac{|\Delta_1|}{\sqrt{\omega^2 + |\Delta_1|^2}} - \frac{|\Delta_1|}{\omega} \right) + \frac{|\Delta_1|}{\Lambda} - |\Delta_1|\ln\frac{T}{T_c} = \frac{\det\begin{pmatrix} |\Delta_1| & \lambda_{12} \\ |\Delta_2| & \lambda_{22} \end{pmatrix}}{\det \lambda_{ij}}, \\ \\ 2\pi T \sum_{\omega>0} \left( \frac{|\Delta_2|}{\sqrt{\omega^2 + |\Delta_2|^2}} - \frac{|\Delta_2|}{\omega} \right) + \frac{|\Delta_2|}{\Lambda} - |\Delta_2|\ln\frac{T}{T_c} = \frac{\det\begin{pmatrix} \lambda_{11} & |\Delta_1| \\ \lambda_{21} & |\Delta_2| \end{pmatrix}}{\det \lambda_{ij}}. \end{cases} \qquad (A.9)$$

In Eqs. (A.6) and (A.7), $D_i$ are the intraband diffusion coefficients, $\Gamma_{ij} = \frac{n_{\text{imp}}}{N_i} \int |u_{\mathbf{k},\mathbf{q}}|^2 \frac{dA_\mathbf{q}}{v_i(\mathbf{q})} \frac{dA_\mathbf{k}}{v_j(\mathbf{q})}$ denote the interband scattering coefficients, $N_i = \int \frac{dA_\mathbf{k}}{v_i(\mathbf{k})}$ corresponds to the density of states at the Fermi level for each zone, $\lambda_{ij} = \frac{1}{N_i} \int \frac{V(\mathbf{k},\mathbf{q})}{v_i(\mathbf{k}) v_j(\mathbf{q})} dA_\mathbf{k} dA_\mathbf{q}$ is the matrix of the constants of the intra- and interband interactions, and $\Lambda$ is the largest eigenvalue of the matrix $\lambda_{ij}$.

For a three-band superconductor, the Usadel equations can be generalized as follows:[167]

$$\omega f_1 - \frac{D_1}{2}\left(g_1\Pi^2 f_1 - f_1\nabla^2 g_1\right) = \Delta_1 g_1 + \Gamma_{12}(g_1 f_2 - g_2 f_1) + \Gamma_{13}(g_1 f_3 - g_3 f_1), \qquad (A.10)$$

$$\omega f_2 - \frac{D_2}{2}\left(g_2\Pi^2 f_2 - f_2\nabla^2 g_2\right) = \Delta_2 g_2 + \Gamma_{21}(g_2 f_1 - g_1 f_2) + \Gamma_{23}(g_2 f_3 - g_3 f_2), \tag{A.11}$$

$$\omega f_3 - \frac{D_3}{2}\left(g_3\Pi^2 f_3 - f_3\nabla^2 g_3\right) = \Delta_3 g_3 + \Gamma_{31}(g_3 f_1 - g_1 f_3) + \Gamma_{32}(g_3 f_2 - g_2 f_3) \tag{A.12}$$

and the self-consistency Eq. (A.8) can be represented in the form

$$\begin{cases} 2\pi T \sum_{\omega>0}\left(\frac{|\Delta_1|}{\sqrt{\omega^2+|\Delta_1|^2}} - \frac{|\Delta_1|}{\omega}\right) + \frac{|\Delta_1|}{\Lambda} - |\Delta_1|\ln\frac{T}{T_c} = \frac{\det\begin{pmatrix} |\Delta_1| & \lambda_{12} & \lambda_{13} \\ |\Delta_2|\cos\phi & \lambda_{22} & \lambda_{23} \\ |\Delta_3|\cos\theta & \lambda_{32} & \lambda_{33} \end{pmatrix}}{\det\lambda_{ij}}, \\[2ex] 2\pi T \sum_{\omega>0}\left(\frac{|\Delta_2|\cos\phi}{\sqrt{\omega^2+|\Delta_2|^2}} - \frac{|\Delta_2|}{\omega}\right) + \frac{|\Delta_2|}{\Lambda} - |\Delta_2|\ln\frac{T}{T_c} = \frac{\det\begin{pmatrix} \lambda_{11} & |\Delta_1| & \lambda_{13} \\ \lambda_{21} & |\Delta_2|\cos\phi & \lambda_{23} \\ \lambda_{31} & |\Delta_3|\cos\theta & \lambda_{33} \end{pmatrix}}{\det\lambda_{ij}}, \\[2ex] 2\pi T \sum_{\omega>0}\left(\frac{|\Delta_3|\cos\theta}{\sqrt{\omega^2+|\Delta_3|^2}} - \frac{|\Delta_3|}{\omega}\right) + \frac{|\Delta_3|}{\Lambda} - |\Delta_3|\ln\frac{T}{T_c} = \frac{\det\begin{pmatrix} \lambda_{11} & \lambda_{12} & |\Delta_1| \\ \lambda_{21} & \lambda_{22} & |\Delta_2|\cos\phi \\ \lambda_{31} & \lambda_{32} & |\Delta_3|\cos\theta \end{pmatrix}}{\det\lambda_{ij}}, \end{cases} \tag{A.13}$$

where, as before, $\Lambda$ is the largest eigenvalue of the matrix $\lambda_{ij}$, and $\varphi$ and $\theta$ denote the phase-difference of the order parameters. The latter can be found from the variation of the energy of a multiband superconductor[113]

$$F = \frac{1}{2}\sum_{ij}\Delta_i\Delta_j^* N_i \lambda_{ij}^{-1} + F_i + F_{\text{imp}}, \tag{A.14}$$

where $\lambda_{ij}^{-1}$ is the inverse matrix of the interaction constants $\lambda_{ij}$ and

$$F_i = 2\pi T \sum_{\omega>0}^{\omega_D} N_i\left[\omega(1-g_i) - \text{Re}\left(f_i^*\Delta_i\right) + \frac{1}{4}D_i\left(\Pi f_i\Pi^* f_i^* + \nabla g_i\nabla g_i\right)\right], \tag{A.15}$$

$$F_{\text{imp}} = 2\pi(N_1\Gamma_{12}+N_2\Gamma_{21})\sum_{\omega>0}^{\omega_D}\left(1-g_1g_2-\text{Re}\left(f_1^*f_2\right)\right) + 2\pi(N_1\Gamma_{13}+N_3\Gamma_{31})\sum_{\omega>0}^{\omega_D}\left(1-g_1g_3-\text{Re}\left(f_1^*f_3\right)\right) \\ +2\pi(N_2\Gamma_{23}+N_3\Gamma_{32})\sum_{\omega>0}^{\omega_D}\left(1-g_2g_3-\text{Re}\left(f_2^*f_3\right)\right). \tag{A.16}$$

Introducing the notation

$$\Omega = \sqrt{1 - \left(\frac{\left(\lambda_{13}^{-1}\right)^2\left(\lambda_{23}^{-1}\right)^2|\Delta_3|^2 - \left(\lambda_{12}^{-1}\right)^2\left(\lambda_{13}^{-1}\right)^2|\Delta_1|^2 - \left(\lambda_{12}^{-1}\right)^2\left(\lambda_{23}^{-1}\right)^2|\Delta_2|^2}{2\left(\lambda_{12}^{-1}\right)^2\lambda_{13}^{-1}\lambda_{23}^{-1}|\Delta_1||\Delta_2|}\right)^2}$$

and assuming the absence of interband impurities for $\phi \in [-\frac{\pi}{2}, \frac{\pi}{2}]$ and $\theta \in [-\frac{\pi}{2}, \frac{\pi}{2}]$ we have

$$\begin{cases} \phi = \pm\arcsin\Omega, \\ \theta = \mp\arcsin\left(\frac{\lambda_{12}^{-1}|\Delta_2|}{\lambda_{13}^{-1}|\Delta_3|}\Omega\right), \end{cases} \tag{A.17}$$

$$\begin{cases} \phi=0, \\ \theta=0, \end{cases} \tag{A.18}$$

for $\phi \in [\frac{\pi}{2}, \frac{3\pi}{2}]$ and $\theta \in [-\frac{\pi}{2}, \frac{\pi}{2}]$

$$\begin{cases} \phi = \pi\pm\arcsin\Omega, \\ \theta=\pm\arcsin\left(\frac{\lambda_{12}^{-1}|\Delta_2|}{\lambda_{13}^{-1}|\Delta_3|}\Omega\right), \end{cases} \tag{A.19}$$

$$\begin{cases} \phi=\pi, \\ \theta=0, \end{cases} \quad (A.20)$$

for $\phi \in [-\frac{\pi}{2}, \frac{\pi}{2}]$ and $\theta \in [\frac{\pi}{2}, \frac{3\pi}{2}]$

$$\begin{cases} \phi = \pm\arcsin\Omega, \\ \theta = \pi \pm \arcsin\left(\frac{\lambda_{12}^{-1}|\Delta_2|}{\lambda_{13}^{-1}|\Delta_3|}\Omega\right), \end{cases} \quad (A.21)$$

$$\begin{cases} \phi=0, \\ \theta=\pi, \end{cases} \quad (A.22)$$

and, finally, for, $\phi \in [\frac{\pi}{2}, \frac{3\pi}{2}]$ and $\theta \in [\frac{\pi}{2}, \frac{3\pi}{2}]$

$$\begin{cases} \phi = \pi \pm \arcsin\Omega, \\ \theta = \pi \mp \arcsin\left(\frac{\lambda_{12}^{-1}|\Delta_2|}{\lambda_{13}^{-1}|\Delta_3|}\Omega\right), \end{cases} \quad (A.23)$$

$$\begin{cases} \phi=\pi, \\ \theta=\pi. \end{cases} \quad (A.24)$$

The choice of the respective solutions (A.17)–(A.24) is determined by a special system of inequalities, which follows from the stability condition for the energy function $F(\phi,\theta)$ [see Eq. (A.14)]

$$\begin{cases} \frac{\partial^2 F}{\partial \phi^2} < 0, \\ \frac{\partial^2 F}{\partial \phi^2} \cdot \frac{\partial^2 F}{\partial \theta^2} - \left(\frac{\partial^2 F}{\partial \phi \partial \theta}\right)^2 > 0. \end{cases} \quad (A.25)$$

### B. Phenomenological approach

In the clean limit, the Ginzburg-Landau functional for a two-band superconductor has the form[168,169]

$$F = \int \left[\alpha_1|\psi_1|^2 + \alpha_2|\psi_2|^2 + \frac{1}{2}\beta_1|\psi_1|^4 + \frac{1}{2}\beta_2|\psi_2|^4 + K_1|\mathbf{\Pi}\psi_1|^2 + K_2|\mathbf{\Pi}\psi_2|^2 - \gamma(\psi_1^*\psi_2 + \psi_2^*\psi_1)\right]dV, \quad (B.1)$$

where

$$\alpha_{1,2} = \left[\frac{\lambda_{22}}{(\lambda_{11}\lambda_{22} - \lambda_{12}\lambda_{21})} - \ln\frac{2\gamma\omega_D}{\pi T}\right]N_{1,2}, \quad \beta_{1,2} = \frac{7\zeta(3)N_{1,2}}{16\pi^2 T_c^2}, \quad \gamma = \frac{\lambda_{12}N_1}{\det\lambda_{ij}} = \frac{\lambda_{21}N_2}{\det\lambda_{ij}}, \quad K_{1,2} = \frac{7\zeta(3)N_{1,2}}{16\pi^2 T_c^2}\langle v_{F_{1,2}}^2\rangle,$$

where $\gamma$ under the natural logarithm is the Euler-Mascheroni constant (not to be confused with the phenomenological coefficient taking into account the interband interaction), and the integration in Eq. (B.1) is carried out over the entire one-, two- or three-dimensional superconductor.

In the dirty limit, for the case of weak interband scattering $\Gamma_{21}, \Gamma_{12} \ll 1$, the Ginzburg-Landau functional becomes more complicated:[170]

$$F = \int \left[\alpha_1|\psi_1|^2 + \alpha_2|\psi_2|^2 + \frac{1}{2}\beta_1|\psi_1|^4 + \frac{1}{2}\beta_2|\psi_2|^4 + K_{1i}|\mathbf{\Pi}\psi_1|^2 + K_{2i}|\mathbf{\Pi}\psi_2|^2 - \gamma(\psi_1^*\psi_2 + \psi_2^*\psi_1)\right.$$
$$\left. + \eta(\mathbf{\Pi}_i\psi_1\mathbf{\Pi}_i^*\psi_2^* + \mathbf{\Pi}_i^*\psi_1^*\mathbf{\Pi}_i\psi_2) - \nu|\psi_1|^2|\psi_2|^2 + 2\nu\left(|\psi_1|^2 + |\psi_2|^2\right)\text{Re}(\psi_1\psi_2)\right]dV, \quad (B.2)$$

where

$$\alpha_1 = \frac{N_1}{2}\left(\ln\frac{T}{T_{c1}} + \frac{\pi\Gamma_{12}}{4T_c}\right), \quad \alpha_2 = \frac{N_2}{2}\left(\ln\frac{T}{T_{c2}} + \frac{\pi\Gamma_{21}}{4T_c}\right), \quad \beta_1 = N_1\left(\frac{7\zeta(3)}{16\pi^2 T_c^2} - \frac{3\pi\Gamma_{12}}{384 T_c^3}\right), \quad \beta_2 = N_2\left(\frac{7\zeta(3)}{16\pi^2 T_c^2} - \frac{3\pi\Gamma_{21}}{384 T_c^3}\right),$$

$$K_1 = N_1 D_1\left(\frac{\pi}{16 T_c} - \frac{7\zeta(3)\Gamma_{12}}{8\pi^2 T_c^2}\right), \quad K_2 = N_2 D_2\left(\frac{\pi}{16 T_c} - \frac{7\zeta(3)\Gamma_{21}}{8\pi^2 T_c^2}\right),$$

Here

$$\gamma = \frac{N_1}{2}\left(\frac{\lambda_{12}}{(\lambda_{11}\lambda_{22} - \lambda_{12}\lambda_{21})} + \frac{\pi\Gamma_{12}}{4T_c}\right) + \frac{N_2}{2}\left(\frac{\lambda_{21}}{(\lambda_{11}\lambda_{22} - \lambda_{12}\lambda_{21})} + \frac{\pi\Gamma_{21}}{4T_c}\right), \quad \eta = \frac{7\zeta(3)}{(4\pi T_c)^2}(D_1 + D_2)(\Gamma_{12}N_1 + \Gamma_{21}N_2),$$

$$\nu = \frac{\pi}{384 T_c^3}(\Gamma_{12}N_1 + \Gamma_{21}N_2).$$

Here

$$T_1 = T_c \exp\left(-\frac{\sqrt{(\lambda_{11} - \lambda_{22})^2 + 4\lambda_{12}\lambda_{21}} - (\lambda_{11} - \lambda_{22})}{2\det\lambda}\right), \quad \text{and} \quad T_2 = T_c \exp\left(-\frac{\sqrt{(\lambda_{11} - \lambda_{22})^2 + 4\lambda_{12}\lambda_{21}} + (\lambda_{11} - \lambda_{22})}{2\det\lambda}\right)$$

represent the critical temperatures of individual, non-interacting zones.

For $\Gamma_{12}, \Gamma_{21} \approx T_c$, the Ginzburg-Landau functional can be reduced to the diagonal form

$$F = \int \left[\alpha_1|\psi_1|^2 + \alpha_2|\psi_2|^2 + \frac{1}{2}\beta_1|\psi_1|^4 + \frac{1}{2}\beta_2|\psi_2|^4 + \hbar^2\left(\frac{|\psi_1|^2}{2m_1} + \frac{|\psi_2|^2}{2m_2} + 2\eta|\psi_1||\psi_2|\cos\phi\right)(\nabla\theta)^2 \right.$$
$$\left. + \hbar^2\left(c_2^2\frac{|\psi_1|^2}{2m_1} + c_1^2\frac{|\psi_2|^2}{2m_2} - 2\eta c_1 c_2 |\psi_1||\psi_2|\cos\phi\right)(\nabla\phi)^2 - 2\gamma|\psi_1||\psi_2|\cos\phi\right]dV. \tag{B.3}$$

Here, $\phi = \varphi_1 - \varphi_2$ is the phase difference of the order parameters, and $\theta = c_1\chi_1 + c_2\chi_2$, where

$$c_1 = \frac{\frac{|\psi_1|^2}{m_1} + 2\eta|\psi_1||\psi_2|\cos\phi}{\frac{|\psi_1|^2}{m_1} + \frac{|\psi_2|^2}{m_2} + 4\eta|\psi_1||\psi_2|\cos\phi}, \quad c_2 = \frac{\frac{|\psi_2|^2}{m_2} + 2\eta|\psi_1||\psi_2|\cos\phi}{\frac{|\psi_1|^2}{m_1} + \frac{|\psi_2|^2}{m_2} + 4\eta|\psi_1||\psi_2|\cos\phi}. \tag{B.4}$$

For arbitrary values of the microscopic coefficients of interband scattering $\Gamma_{12}$ and $\Gamma_{21}$, the Ginzburg-Landau functional in a uniform current-free state can be written as[171]

$$F = \int \left[a_{11}|\psi_1|^2 + a_{22}|\psi_2|^2 + \frac{1}{2}b_{11}|\psi_1|^4 + \frac{1}{2}b_{22}|\psi_2|^4 + 2a_{12}|\psi_1||\psi_2|\cos\phi + b_{12}|\psi_1|^2|\psi_2|^2 \right.$$
$$\left. + 2\left(c_{11}|\psi_1|^2|\psi_2| + c_{22}|\psi_1||\psi_2|^2\right)\cos\phi + c_{12}|\psi_1|^2|\psi_2|^2\cos 2\phi\right]dV, \tag{B.5}$$

where $\phi$ is the phase difference of the order parameter, and the phenomenological coefficients are expressed through microscopic parameters:

$$a_{ii} = N_i\left(\frac{\lambda_{ii}}{\det\lambda_{ij}} - 2\pi T\sum_{\omega>0}^{\omega_D}\frac{\omega + \Gamma_{ij}}{\omega(\omega + \Gamma_{ij} + \Gamma_{ji})}\right), \quad a_{ij} = -N_i\left(\frac{\lambda_{ij}}{\det\lambda_{ij}} + 2\pi T\sum_{\omega>0}^{\omega_D}\frac{\Gamma_{ij}}{\omega(\omega + \Gamma_{ij} + \Gamma_{ji})}\right),$$

$$b_{ii} = N_i\pi T\sum_{\omega>0}^{\omega_D}\frac{(\omega + \Gamma_{ij})^4}{\omega^3(\omega + \Gamma_{ij} + \Gamma_{ji})^4} + N_i\pi T\sum_{\omega>0}^{\omega_D}\frac{\Gamma_{ij}(\omega + \Gamma_{ji})(\omega^2 + 3\omega\Gamma_{ji} + \Gamma_{ij}^2)}{\omega^3(\omega + \Gamma_{ij} + \Gamma_{ji})^4},$$

$$b_{ij} = -N_i\pi T\sum_{\omega>0}^{\omega_D}\frac{\Gamma_{ij}}{\omega^3(\omega + \Gamma_{ij} + \Gamma_{ji})^4} + N_i\pi T\sum_{\omega>0}^{\omega_D}\frac{\Gamma_{ij}(\Gamma_{ij} + \Gamma_{ji})(\Gamma_{ji}(\omega + 2\Gamma_{ij}) + \omega\Gamma_{ij})}{\omega^3(\omega + \Gamma_{ij} + \Gamma_{ji})^4},$$

$$c_{ii} = N_i\pi T\sum_{\omega>0}^{\omega_D}\frac{\Gamma_{ij}(\omega + \Gamma_{ji})(\omega^2 + \omega(\omega + \Gamma_{ji})(\Gamma_{ij} + \Gamma_{ji}))}{\omega^3(\omega + \Gamma_{ij} + \Gamma_{ji})^4}, \quad c_{ij} = N_i\pi T\sum_{\omega>0}^{\omega_D}\frac{\Gamma_{ij}(\omega + \Gamma_{ij})(\omega + \Gamma_{ji})(\Gamma_{ij} + \Gamma_{ji})}{\omega^3(\omega + \Gamma_{ij} + \Gamma_{ji})^4}.$$

A three-band superconductor in the clean limit and also in the current-free state can be described by the following Ginzburg-Landau functional[167,172–177]

$$F = \int \left[\alpha_1|\psi_1|^2 + \alpha_2|\psi_2|^2 + \alpha_3|\psi_3|^2 + \frac{1}{2}\beta_1|\psi_1|^4 + \frac{1}{2}\beta_2|\psi_2|^4 + \frac{1}{2}\beta_3|\psi_3|^4 + K_{1i}|\mathbf{\Pi}\psi_1|^2 + K_{2i}|\mathbf{\Pi}\psi_2|^2 \right.$$
$$\left. + K_{3i}|\mathbf{\Pi}\psi_3|^2 - \gamma_{12}\left(\psi_1^*\psi_2 + \psi_2^*\psi_1\right) - \gamma_{13}\left(\psi_1^*\psi_3 + \psi_1\psi_3^*\right) - \gamma_{23}\left(\psi_2^*\psi_3 + \psi_2\psi_3^*\right)\right]dV, \tag{B.6}$$

where

$$\alpha_1 = \left(\ln\left(\frac{2\gamma\langle\omega_0\rangle}{\pi T}\right) - \frac{\lambda_{22}\lambda_{33} - \lambda_{23}\lambda_{32}}{\det(\lambda)}\right)N_1, \quad \alpha_2 = \left(\ln\left(\frac{2\gamma\langle\omega_0\rangle}{\pi T}\right) - \frac{\lambda_{11}\lambda_{33} - \lambda_{13}\lambda_{31}}{\det(\lambda)}\right)N_2,$$

$$\alpha_3 = \left(\ln\left(\frac{2\gamma\langle\omega_0\rangle}{\pi T}\right) - \frac{\lambda_{11}\lambda_{22} - \lambda_{12}\lambda_{21}}{\det(\lambda)}\right)N_3, \quad K_i = \frac{\pi D_i N_i}{8 T_c}, \quad \beta_i = \frac{7\zeta(3)N_i}{8\pi^2 T_c^2}, \quad \gamma_{12} = \frac{(\lambda_{12}\lambda_{33} - \lambda_{13}\lambda_{32})N_2}{\det(\lambda)},$$

$$\gamma_{13} = \frac{(\lambda_{13}\lambda_{22} - \lambda_{12}\lambda_{23})N_3}{\det(\lambda)}, \quad \gamma_{23} = \frac{(\lambda_{11}\lambda_{23} - \lambda_{13}\lambda_{21})N_3}{\det(\lambda)}.$$

In the dirty limit, in the absence of external currents, the free energy is expressed as[178]

$$\begin{aligned}F = \int &\left[\alpha_1|\psi_1|^2 + \alpha_2|\psi_2|^2 + \alpha_3|\psi_3|^2 + \frac{1}{2}\beta_1|\psi_1|^4 + \frac{1}{2}\beta_2|\psi_2|^4 + \frac{1}{2}\beta_3|\psi_3|^4 + K_{1i}|\mathbf{\Pi}\psi_i|^2 + K_{2i}|\mathbf{\Pi}\psi_2|^2 + K_{3i}|\mathbf{\Pi}\psi_3|^2\right.\\ &-\gamma_{12}\left(\psi_1^*\psi_2 + \psi_2^*\psi_1\right) - \gamma_{13}\left(\psi_1^*\psi_3 + \psi_1\psi_3^*\right) - \gamma_{23}\left(\psi_2^*\psi_3 + \psi_2\psi_3^*\right) + b_{12}|\psi_1|^2|\psi_2|^2 + b_{13}|\psi_1|^2|\psi_3|^2 + b_{23}|\psi_2|^2|\psi_3|^2\\ &\times c_{11}\left(|\psi_1|^3|\psi_2| + |\psi_1|^3|\psi_3|\right)\cos(\varphi_1 - \varphi_2) + c_{22}\left(|\psi_2|^3|\psi_1| + |\psi_2|^3|\psi_3|\right)\cos(\varphi_1 - \varphi_3) + c_{33}\left(|\psi_3|^3|\psi_1| + |\psi_3|^3|\psi_2|\right)\\ &\times \cos(\varphi_2 - \varphi_3) \times c_{12}|\psi_1|^2|\psi_2|^2\cos 2(\varphi_1 - \varphi_2) + c_{13}|\psi_1|^2|\psi_3|^2\cos 2(\varphi_1 - \varphi_3) + c_{23}|\psi_2|^2|\psi_3|^2\cos 2(\varphi_2 - \varphi_3)\\ &+d_{123}|\psi_1|^2|\psi_2||\psi_3|\cos(2\varphi_1 - \varphi_2 - \varphi_3) + d_{231}|\psi_2|^2|\psi_3||\psi_1|\cos(2\varphi_2 - \varphi_1 - \varphi_3) + d_{312}|\psi_3|^2|\psi_1||\psi_2|\cos(2\varphi_3 - \varphi_1 - \varphi_2).\end{aligned}$$
(B.7)

The coefficients in expression (B7) also have a microscopic representation.

By varying the above free-energy functionals in $\psi_1$, $\psi_2$ and $\mathbf{A}$, the corresponding Ginzburg-Landau equations describing a multiband superconductor in the clean or dirty limit can be obtained.

a)Email: yuriyyerin@gmail.com